\documentclass[10pt]{iopart}

\usepackage{iopams}
\expandafter\let\csname equation*\endcsname\relax
\expandafter\let\csname endequation*\endcsname\relax
\usepackage{amsmath}
\usepackage{graphicx}
\usepackage{subfigure}
\usepackage{geometry}
\usepackage{hyperref}
\usepackage{color}
\usepackage{float}
 
\begin{document}

\topical{Spatio-temporal characterization of ultrashort laser beams: a tutorial}
\author{Spencer W. Jolly, Olivier Gobert, and Fabien Qu{\'e}r{\'e}}
\address{LIDYL, CEA, CNRS, Universit{\'e} Paris-Saclay, CEA Saclay, 91 191 Gif-sur-Yvette, France}
\eads{\mailto{spencer.jolly@cea.fr}, \mailto{fabien.quere@cea.fr}}

\vspace{10pt}
\begin{indented}
\item[]originally from 2 March 2020, on arXiv 8 July 2020
\end{indented}

\begin{abstract}
The temporal characterization of ultrafast laser pulses has become a cornerstone capability of ultrafast optics laboratories and is routine both for optimizing laser pulse duration and designing custom fields. Beyond pure temporal characterization, spatio-temporal characterization provides a more complete measurement of the spatially-varying temporal properties of a laser pulse. These so-called spatio-temporal couplings (STCs) are generally nonseparable chromatic aberrations that can be induced by very common optical elements -- for example diffraction gratings and thick lenses or prisms made from dispersive material. In this tutorial we introduce STCs and a detailed understanding of their behavior in order to have a background knowledge, but also to inform the design of characterization devices. We then overview a broad range of spatio-temporal characterization techniques with a view to mention most techniques, but also to provide greater details on a few chosen methods. The goal is to provide a reference and a comparison of various techniques for newcomers to the field. Lastly, we discuss nuances of analysis and visualization of spatio-temporal data, which is an often underappreciated and non-trivial part of ultrafast pulse characterization.
\end{abstract}

%
%
%
 
\ioptwocol

\section{Introduction}
\label{sec:intro}


The frequency dependence of the spatial properties of a broadband light beam or of the optical response of a system is known as chromatism, and has been discussed for decades in many different fields of classical optics. In photography for example, chromatism of the imaging lens affects the ability to properly image an object illuminated by ambient white light, because slightly different images are produced for each color of the incident light. 

Due to the time-frequency uncertainty principle, ultrashort laser beams necessarily have significant spectral widths, and can therefore also be affected by chromatism. As for any other broadband light source, this impacts the spatial properties of the beam: if a chromatic ultrashort laser beam is focused by a perfect optic, its different frequency components are focused differently, resulting in a degradation of the spatial concentration of the laser light at focus.

Yet, compared to incoherent broadband light, chromatism has further consequences for this peculiar type of light sources, now in the time domain: if the spectral properties (in amplitude and phase) of the laser beam are position-dependent, then by Fourier-transformation its temporal properties vary in space too. Such a dependence is known as a spatio-temporal coupling (STC), and implies that chromatism not only affects the concentration of light energy in space, but also its bunching in time, which is the key feature of ultrashort lasers. Properly assessing the impact of chromatism on ultrashort lasers therefore requires specific measurement methods, which give access to the full \textsl{spatio-temporal} structure of these beams.

Developing such a spatio-temporal metrology, up to the point where it becomes part of the standard characterization routine of ultrashort lasers, is essential because STCs can have highly detrimental effects on the performance of these lasers. As is clear from previous qualitative analyses, they often have the effect of increasing the pulse duration and reducing intensity in focus~\cite{bourassin-bouchet11}, but can also have more complex yet very relevant effects, for example on pulse contrast~\cite{li17,li18-1}. On the other hand, STCs also provide extremely powerful ways of controlling the properties of light beams and therefore laser-matter interaction processes. Examples include optimization of non-colinear sum- or difference-frequency generation~\cite{martinez89,maznev98,huangS-W12,gobert14}), broadband THz generation~\cite{stepanov03,fulop14}, isolated attosecond pulse generation by the attosecond lighthouse effect~\cite{vincenti12,wheeler12,kim13,quere14,auguste16}, improved non-linear microscopy using spatio-temporal focusing~\cite{DURST20081796}, and even laser machining~\cite{sun18,wangP18,liQ19}. 

There is a broad collection of purely temporal laser diagnostics~\cite{stibenz06,walmsley09}, which are meant to characterize the evolution of the electric field of a laser pulse in time. These measurements are generally either an average over a given aperture of the pulse, or essentially done at a single point (i.e. a small aperture), and therefore the result is only the local electric field resolved in time. These techniques include frequency-resolved optical gating (FROG)~\cite{kane93,trebino97,oshea01,bates10}, spectral phase interferometry for direct electric-field reconstruction (SPIDER)~\cite{iaconis98,gallmann99,mairesse05,radunsky07,mahieu15}, self-referenced spectral interferometry (SRSI, WIZZLER device)~\cite{oksenhendler10,moulet10,trisorio12,oksenhendler12}, and D-Scan~\cite{miranda11,loriot13} among others. The devices and techniques to characterize a laser pulse spatio-temporally are often related to these purely temporal techniques, but also can employ completely separate schemes. Although not a pre-requisite, prior knowledge of temporal measurement techniques for ultrashort pulses will facilitate the reading of this tutorial. Extensive reviews, tutorials or even courses can be found in various past works~\cite{Monmayrant_2010,dorrer19}. 

This tutorial aims not to review the entire field of spatio-temporal metrology, especially since there has been an extremely comprehensive review done very recently~\cite{dorrer19}. In contrast, it aims to introduce spatio-temporal couplings and a large range of techniques to diagnose them, in a manner to guide those without significant experience on this topic. We hope that scientists can use this tutorial to determine how to most simply and correctly diagnose or control spatio-temporal couplings in their specific situation.

Section~\ref{sec:concepts} is mostly devoted to defining STCs in a pedagogical way, and introducing the characteristics of the most basic and common couplings. We finish this section by first touching upon techniques that require a minimal amount of specialized equipment, but may not be able to measure arbitrary STCs. In sections~\ref{sec:spatial} and \ref{sec:frequency} we will then expand to more complete and advanced techniques, which are intended to determine the complete spatio-temporal structure of ultrashort laser beams. This ideally requires sampling a field in a three-dimensional space (two spatial coordinates, and time or frequency). This can be considered as one of the main difficulties of STC metrology, since the main light sensors available to date are cameras, which only have two dimensions. This problem has often been circumvented by resolving one spatial dimension only, obviously at the cost of a significant and potentially highly detrimental loss of information. Many present techniques are actually affected by this limitation, but will nonetheless be discussed in this tutorial due to their importance in the development of this field.

Spatio-temporal or spatio-spectral metrology uses in general one of two methodologies: resolving a complete temporal or spectral characterization method in one (or more) spatial dimension(s) ('spatially-resolved spectral measurements'), or resolving the amplitude and phase of a spatial measurement at multiple frequencies ('frequency-resolved spatial measurements'). Although the separation based on these definitions can sometimes be difficult to distinguish, the two sections on 'complete' techniques will be delineated according to our interpretation of these descriptions.

The outcome of a complete measurement is a three-dimensional complex matrix describing the $E$-field of the laser beam in space-time or space-frequency. Interpreting and exploiting such a measurement result is far from straightforward, and the visualization and analysis of such datasets can therefore be considered as another significant difficulty of STC metrology. Specific tools have been developed over the last few years, and are summarized in the final section of this tutorial. 

\section{Key concepts of spatio-temporal couplings and their metrology}
\label{sec:concepts}

Before discussing specific advanced methods to characterize the spatio-temporal properties of ultrashort laser pulses, it is necessary to understand exactly what STCs are, the implications on the beam properties in different parameter spaces, and the first very simple steps one might take to diagnose the presence of STCs, at least qualitatively. This is necessary to understand the capabilities of a given measurement device, i.e. it is crucial to understand what forms low- or high-order STCs may take at the measurement position. This is also helpful to finally analyze the result of any complete or incomplete measurement.

The goal of any characterization device is to measure as completely as possible the 3-dimensional electric field of an ultrashort laser pulse $E$ in space and time $E(x,y,t)$, or in space and frequency $\hat{E}(x,y,\omega)$ (for the sake of simplicity, we will assume throughout this paper that the field is linearly-polarized, with the same polarization direction all across the beam). The quantities $E$ and $\hat{E}$ are related to each other by the one dimensional Fourier transform from time to frequency. We use $x$ and $y$ as the transverse dimensions, where the beam is propagating along $z$. Because a fully-characterized beam can be numerically propagated to any $z$, we are interested in the measurement of $E$ at only one $z$ that depends on the characterization device in use. 

In each case the field is composed of an amplitude term and a phase term, i.e.: $E(x,y,t)=\sqrt{I(x,y,t)}e^{i\phi(x,y,t)}$ and $\hat{E}(x,y,\omega)=\sqrt{\hat{I}(x,y,\omega)}e^{i\hat{\phi}(x,y,\omega)}$, where we will sometimes refer to the intensity $\hat{I}(x,y,\omega)$ or the amplitude $\hat{A}(x,y,\omega)=\sqrt{\hat{I}(x,y,\omega)}$. We will use these notations for the rest of the tutorial.

The function $\hat{\phi}(x,y,\omega)$ is the 'spatio-spectral phase', a crucial quantity for the properties of ultrashort laser beams. Much of the complexity of understanding and measuring STCs is actually concentrated in this function. It is closely related to the simple spectral phase $\hat{\phi}(\omega)$ provided by usual temporal measurement devices, in that $\hat{\phi}(\omega)$ is either the value of $\hat{\phi}(x,y,\omega)$ at a test position $(x_0, y_0)$, or a spatial average of this function over $x$ and $y$. Just as in the case of temporal metrology, it is a much simpler problem to measure only the spatio-spectral amplitude or intensity, but measuring both the amplitude and phase is more challenging and will be the topic of the two next sections of this tutorial.

We feel that it is important to note finally that the term 'spatio-temporal' and other similar versions of the term are often used to refer to the combination of measurements that are simply spatial and temporal. This has sometimes been the case in pulse characterization, but is much more often the case in fields that are more far-afield such as microscopy or spectroscopy, where it is less common to also have temporal information in the first place. Our definition is much stronger, i.e. in this tutorial a spatio-temporal measurement is not just the addition of a spatial measurement device and a temporal measurement device, but it is the measurement of the full spatio-temporal field (whether there are STCs present or not).

\subsection{The general concept of spatio-temporal coupling}
\label{sec:concepts_general}

For the purposes of this report we define the basic concept of what a spatio-temporal coupling actually is, in the most simple terms possible. That is: a spatio-temporal coupling is any property of an ultrashort laser pulse that results in the inability to describe the electric field of the laser pulse as a product of functions in space and time. Mathematically, if a beam has STCs, then the following statement is true:

\begin{equation}
\label{eq:STC}
E(x,y,t)\neq f(x,y)\times g(t) \quad \forall \quad f(x,y), g(t) .
\end{equation}

\noindent In such a case, a similar inequality holds for $\hat{E}(x,y,\omega)$, since it is related to $E(x,y,t)$ by a simple Fourier transform with respect to time. In other words, as mentioned in the introduction, a beam with spatio-temporal couplings also has spatio-spectral couplings (i.e. chromatism), and we will often use these terms interchangeably. In fact the representation in frequency space is often the more convenient one to analyze the beam properties.

\begin{figure}[htb]
	\centering
	\includegraphics[width=83mm]{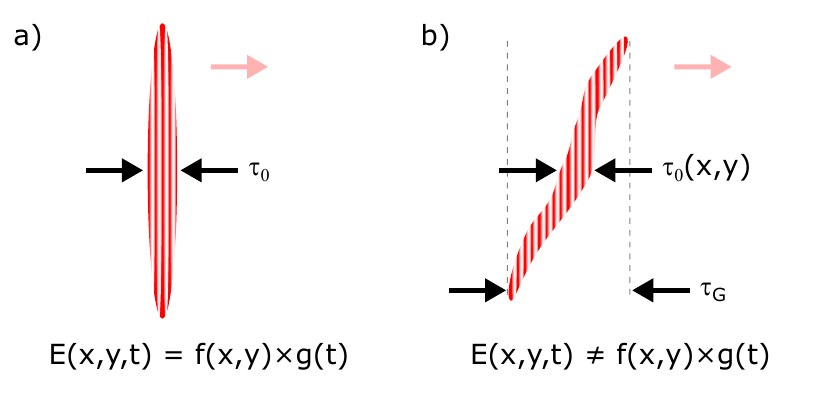}
	\caption{Basic concept of STCs. Both panels show a sketch of the spatio-temporal electric field of an ultrashort laser beam. In (a), this a beam without STCs, where the full electric field can be expressed by separable functions, and the local pulse duration $\tau_0$ is valid globally. In (b), the beam has significant STCs, where the field is no longer separable and the local duration $\tau_0$ is different than the global duration $\tau_G$. The carrier wave here is a sketch and not meant to be to scale.}
	\label{fig:STC_concept}
\end{figure}

An example of a nonseparable beam can be seen in a sketch in Fig.~\ref{fig:STC_concept}, where the beam in panel (a) has no STCs, and the beam in panel (b) does. The example with no STCs is perfectly described by separable functions $f(x,y)$ and $g(t)$ in space and time respectively. For the example in Fig.~\ref{fig:STC_concept}(b) there is both a varying arrival time of the pulse with the transverse dimension and some transverse variation in the temporal width. To account for the former effect, one may naively describe the field now in terms of $g(t-\tau_0(x))$ with $\gamma$ according to the magnitude of the tilt of the pulse. This is quite simple and potentially valid, but would still result in the full field $E(x,y,t)$ no longer being separable.

This distinction is simple to see when the mathematical descriptions of the fields are compared. We consider a Gaussian beam in space and time for convenience. If $r^2=x^2+y^2$ and the beam has a spatial width $w$, temporal width $\tau_0$, and central frequency $\omega_0$, then the case of Fig.~\ref{fig:STC_concept}(a) is written simply as 

\begin{equation}
\label{eq:GaussNoSTC}
E_{1a}\propto e^{-r^2/w^2}e^{-t^2/\tau_0^2}e^{i\omega_0 t} .
\end{equation}

\noindent This case is clearly separable. If the pulse has the properties shown in Fig.~\ref{fig:STC_concept}(b), then the field is written as

\begin{equation}
\label{eq:GaussYesSTC}
E_{1b}\propto e^{-r^2/w^2}e^{-(t-\tau_0(x))^2/\tau(r)^2}e^{i\omega_0 t} .
\end{equation}

\noindent This is non-separable. As mentioned, this non-separability also has implications on the description of the electric field in frequency and space, but it then takes a different specific form, as will be further discussed in Section~\ref{sec:concepts_manifestations}.

Beyond having an impact on the mathematical description of the electric field, the presence of an STC will also affect measurable parameters. The most obvious is the temporal duration, which in the presence of some STCs could have spatial variation. It is not in the case of all STCs that the local duration will vary in space, but it is true that with any STC there will be a difference between the local pulse duration and the global pulse duration~\cite{bourassin-bouchet11}, referred to as $\tau_0$ and $\tau_G$ respectively in Fig.~\ref{fig:STC_concept}. This generally results in a decrease in the peak intensity, and sometimes a varied spatial distribution of the different frequencies within the beam. The next few sections will discuss the nuances of the previous statements and classify a few of the well-known STCs.

\begin{figure*}[ht]
	\centering
	\includegraphics[width=171mm]{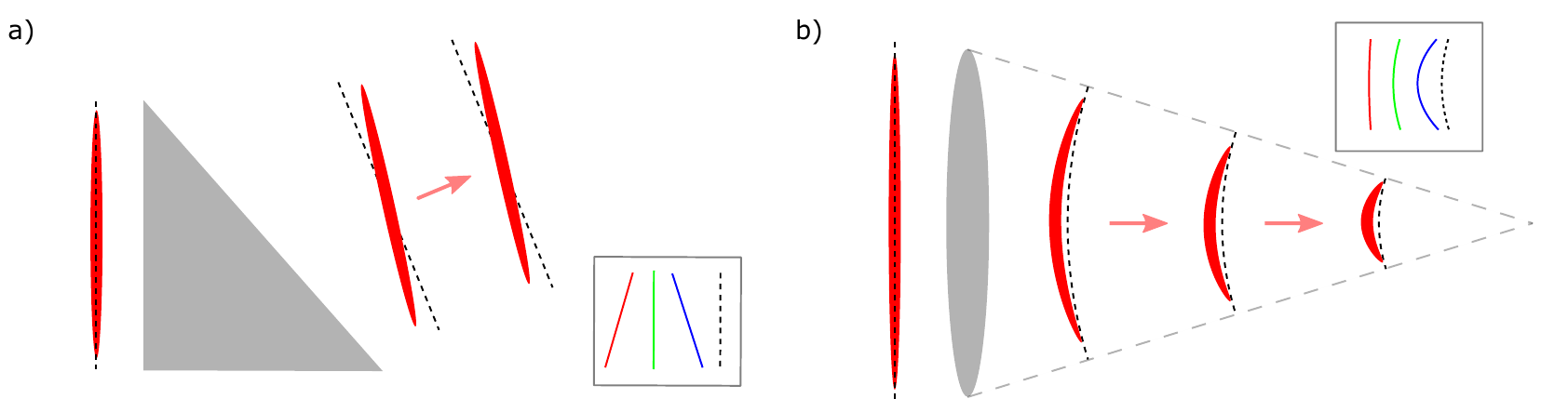}
	\caption{Qualitative examples of simple couplings and their origins. In (a) pulse-front tilt is produced via propagation in a wedged dispersive material. In (b) pulse-front curvature is induced by propagation in a chromatic singlet lens made of a dispersive material. In both cases the inset is showing the phase fronts at three frequencies, with the frequency-average phase front as the dashed line. The sketch in (b) is inspired by and very similar to that in Ref.~\cite{bor89-1}.}
	\label{fig:STC_simple}
\end{figure*}

\subsection{Introduction to low-order couplings}
\label{sec:concepts_low-order}

Here we introduce some common-place low-order STCs, which provide highly instructive examples. The term low-order refers to STCs where the field variations in space-time or space-frequency can be described by low-order polynomials of position coordinates and time/frequency, and that are therefore more likely to occur. For more complete analysis, we urge the reader to reference significant past work on describing and reviewing this topic~\cite{akturk05,gabolde07,akturk10}, and work that has gone over alternative matrix-based formalism specifically designed to describe dispersive optical systems~\cite{kostenbauder90,lin95,marcus16}.

The most prevalent and lowest-order STC is pulse-front tilt (PFT), where the duration of the beam is constant in space, but the arrival time varies linearly with one spatial dimension (Fig.~\ref{fig:STC_simple}(a)). With PFT, both the wavefront and the pulse-front (describing the location of the electromagnetic energy, i.e. the pulse envelope) of the beam are perfectly flat, but they are constantly at an angle to each other. In other words, the pulse front is tilted with respect to the propagation direction of the pulse. The next most common STC is pulse-front curvature (PFC), where the duration is still constant in space, but the arrival time now varies quadratically with the radial position (Fig.~\ref{fig:STC_simple}(b)).

These two canonical STCs, PFT and PFC, can be caused by very simple and commonplace optical elements. For example, PFT can be induced via propagation through a wedged prism of any dispersive optical material (which includes glasses, the most ubiquitous optical materials), as seen in Fig.~\ref{fig:STC_simple}(a). The portion of the beam traveling though the thin part of the prism has traversed less material, so then the accumulated group-delay is less than that of the part of the beam passing through the thick portion of the prism. Because the thickness of the prism linearly depends on the transverse dimension in the plane of the page, then the accumulated group-delay will depend linearly on position as well, and this results in the rotation of the pulse front after propagation through the prism. For the same reason, the phase fronts also get tilted, but they do so according to the phase refractive index rather than the group refractive index. The output beam will have PFT if these two rotation angles are different, which occurs if the phase velocity $v_p$ is different than the group velocity $v_g$, i.e. the medium is dispersive. At higher orders, the output beam can also exhibit a spatially-dependent spectral chirp due to the different encountered thicknesses of glass. This generally has negligible impact compared to the PFT.  

PFC has a similar commonplace source, which is simple chromatic singlet lenses~\cite{bor88,bor89-1}, as seen in Fig.~\ref{fig:STC_simple}(b). From the temporal point of view the portion of the beam at the outer edge of the lens will accrue less group delay than the center of the beam, resulting in a radially-varying arrival time. If the medium is dispersive ($v_p\neq v_g$) then the curvature of the pulse-front will be different than that for the phase-front after such a lens. 


\subsection{Manifestations of couplings}
\label{sec:concepts_manifestations}

This section focuses on expanding the descriptions from the previous sections to be more quantitative and to describe couplings in different domains that are relevant for characterization devices and experiments. The main domains we will consider are the near-field (NF) where the beam is collimated (e.g. the output of a laser system), and the far-field (FF) where the beam is at a focus (e.g. where experiments are generally performed), both in time and frequency. The NF and FF of course have broader definitions in classical optics, but for simplicity we will refer to only these two planes as NF and FF throughout this tutorial.

These NF and FF spaces defined in this way are related via the principles of Fourier optics, so that the NF is related to the FF by a two dimensional spatial Fourier transform and a coordinate change depending on the focal length ($x_\textrm{FF}=k_x \lambda f/2\pi$)~\cite{doi:10.1002/0471213748.ch4}. Therefore the FF is technically equivalent to the $(k_x,k_y)$ reciprocal space at the NF plane (see Refs.~\cite{akturk05,akturk10}). However, from the authors' point of view, considering different physical planes separated by propagation and focusing enables a simpler understanding and is physically more relevant than analyzing the field in different mathematical spaces. We must note that the previously mentioned Gaussian generalization~\cite{akturk05,akturk10} also provides important insight into the behavior of different couplings in different domains, and previous work has also gone into great detail specifically on the effect of couplings on the pulse duration in focus~\cite{bourassin-bouchet11}.

A beam that has no STCs can be described in a simple fashion in all four of the relevant domains (NF and FF, both in time and frequency). A beam with a Gaussian spatial distribution, flat wavefront, and a Gaussian temporal envelope in the NF will have the same temporal envelope in the FF and a Gaussian spatial distribution with a waist determined by the focusing conditions. The description in frequency will be similarly straightforward, regardless of if there is non-zero spectral phase. This is essentially the propagated and/or Fourier-transformed results of Eq.~(\ref{eq:GaussNoSTC}), where the Gaussian nature allows for analytical representations in all cases. 

\begin{figure*}[htb]
	\centering
	\includegraphics[width=171mm]{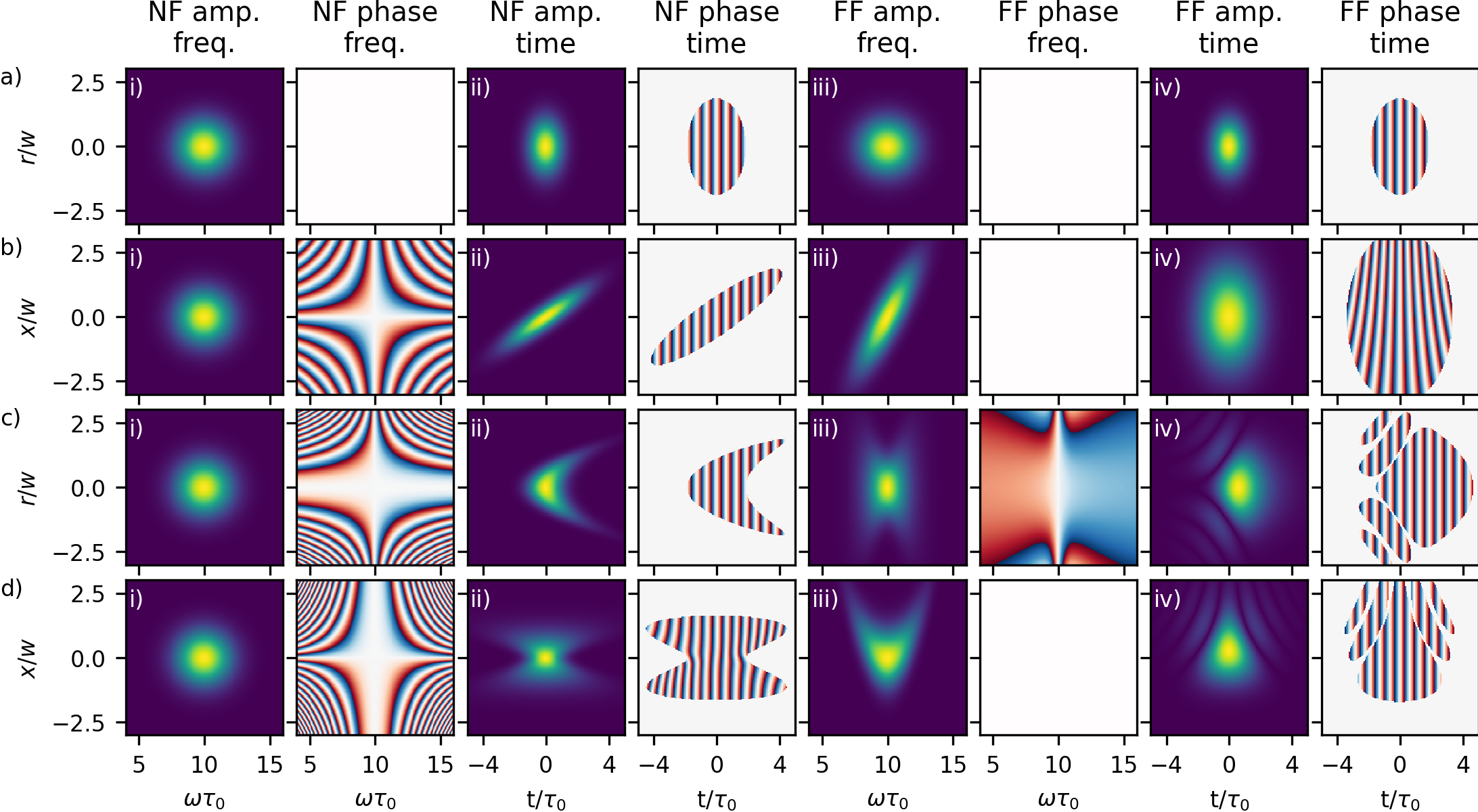}
	\caption{Four quantitative examples of beams, with no STC (a), with AD/PFT (b), with CC/PFC (c), and with a linear chirp varying linearly with radius (d). All the couplings are applied to the beam in the NF. Amplitude and phase are shown for all four examples in the four relevant parameter spaces: both near-field and far-field (collimated beam and in-focus) in both frequency and time. All four examples have the exact same spatio-spectral amplitude in the near-field (sub-panels (i) in each case), so it is therefore the spatio-spectral phase in the near-field that differentiates them. The color scale for amplitude is normalized for each panel and the color scale for phase goes from $-2\pi$ to $2\pi$.}
	\label{fig:all_couplings}
\end{figure*}

We will take this example of a Gaussian STC-free beam as the baseline, which is pictured in all four domains, with amplitude and phase, in the top row (denoted with (a)) of Fig.~\ref{fig:all_couplings}. The NF in frequency is in the pair of panels (i), the NF in time is in panels (ii), the FF in frequency is in (iii), and the FF in time is in (iv). We choose to describe a beam having a Fourier-limited duration $\tau_0$ and central wavelength $\omega_0$ such that $\omega_0\tau_0=10$ (i.e. the beam technically must be very broadband, but we do this to be able to visualize the carrier frequency in time). We plot in normalized units in order to ease the visualization, where for simplicity the NF and FF have a characteristic width noted as $w$ in both cases. In this section we will describe mostly the canonical couplings of PFT and PFC as well as one example beyond that, but we must stress that the final goal of any characterization device is to measure arbitrary couplings. Therefore the importance of this section is to develop the knowledge of how couplings manifest themselve in different spaces.

The first very important intuition is for comparing the properties of a beam in time and in frequency domains, related by a 1D Fourier transform. To this end, in the cases of PFT (Fig.~\ref{fig:all_couplings}(b)(i)) and PFC (Fig.~\ref{fig:all_couplings}(c)(i)), we can first use a simple physical analysis of the optical systems that induce these couplings, before turning to a more formal description. As discussed in the previous section, PFT can be induced by a prism made of a dispersive glass. As is well-known, such a prism induces angular dispersion (AD), i.e. it results in different propagation directions at the prism output for the different frequency components. We can therefore expect PFT (time-domain description) to be  equivalent to a frequency-dependent wavefront tilt (frequency-domain description).

Similarly, PFC can be induced by a chromatic lens, which is known to induce a different wavefront curvature for the different frequency components (CC for chromatic curvature). We can therefore expect PFC (time-domain description) to be  equivalent to a frequency-dependent wavefront curvature (frequency-domain description). We insist on the fact that PFT and AD (or similarly PFC and CC) correspond to the description of the very same beam, but considered in different spaces. Because of this equivalence, these canonical STCs will be referred to as AD/PFT and CC/PFC in the rest of this work.

These correspondences between the time- and frequency-domain descriptions can of course be derived mathematically. A general derivation for arbitrary pulse-front distortions is provided in ~\ref{sec:appendixA} of this tutorial. This simple calculation shows that when considered in the frequency domain, a beam with AD/PFT is characterized by a spatio-spectral phase $\hat{\phi}(x,y,\omega)=\gamma x (\omega-\omega_0)$, where $\gamma$ represents the magnitude of the AD/PFT. This phase is plotted in Fig.~\ref{fig:all_couplings}(b)(i) and can be understood in two ways. One the one hand, this can be considered as a phase varying linearly in frequency, with a slope $\partial \hat{\phi}/\partial \omega$ (corresponding to a delay in the time domain) that varies linearly with position: this describes PFT. On the other hand, this can be considered as a phase varying linearly in position (i.e. a wavefront tilt), with a slope that varies linearly with frequency: this describes AD. Similarly, PFC is described by a  spatio-spectral phase $\hat{\phi}(r,\omega)=\alpha r^2 (\omega-\omega_0)$, where $\alpha$ represents the magnitude of the CC/PFC. This is plotted in Fig.~\ref{fig:all_couplings}(c)(i), and can either be considered as a linear spectral phase with a slope varying quadratically with position (PFC), or as a quadratic spatial phase (wavefront curvature) varying linearly with frequency (CC). 

We have now emphasized multiple times that for a beam with STC, different frequencies have different spatial properties, and this has been nicely illustrated by the previous discussion on AD/PFT and CC/PFC. As result, when a beam affected by chromatism propagates, the different frequency components evolve differently. The beam's spatio-spectral properties and spatio-temporal properties therefore change upon propagation. We now illustrate this point by considering the FF properties of beams that initially have AD/PFT and CC/PFC in the NF. 

To this end, we start from the frequency-domain description of these beams in the NF. For a beam with AD/PFT, the different frequencies have different wavefront tilt. Therefore, in the FF they must have a varying best-focus position in the transverse dimension. This is displayed in Fig.~\ref{fig:all_couplings}(b)(iii), and is known as 'transverse spatial chirp'. As a result of the transformation of the beam upon propagation, the temporal structure of the beam in the FF is also very different from that in the NF. In time, the pulse at focus no longer has any PFT, but has a longer local duration corresponding to the global duration in the NF. The focal spot is spatially larger than that of the perfect reference beam, since different frequencies are focused at different transverse positions. Finally, the spatio-temporal phase has a peculiar structure referred to as wavefront rotation (see the phase map of Fig.~\ref{fig:all_couplings}(b)(iv)). At negative times the spatial phase is tilted in one direction, and over time it changes to finally tilt in the opposite direction at positive times. This describes the fact that the propagation direction of light rotates in time on the scale of the pulse temporal envelope.  This effect has interesting applications in high-intensity optics~\cite{quere14}, in particular for the generation of isolated attosecond pulses.



For a beam with CC/PFC, the different frequencies have a different wavefront curvature in the NF. Therefore, in the FF they must have a varying best-focus position, now along the longitudinal dimension. At a single longitudinal position this manifests as a varying beam size according to frequency, and a spatio-spectral phase that represents the Guoy phase for each frequency. This is seen in panel (c)(iii) of Fig.~\ref{fig:all_couplings}. The pulse in time at focus has a more complex amplitude profile, with a longer duration on-axis and a duration and arrival time that vary with the radial coordinate.

Beyond the visualization just presented in Fig.~\ref{fig:all_couplings}, which had a flat spectral phase for at least one position in the beam, it could be such that a beam with either AD or CC were significantly chirped everywhere in space. This would not change much the spatio-spectral picture, since this simply corresponds to the addition of a spatially-homogeneous spectral phase, but would drastically change the picture in time. Because of this, pulse-front tilt and pulse-front curvature are only strictly proper names for these two couplings with no global chirp, and therefore angular dispersion or chromatic curvature are in a sense more general terms.


To illustrate more complex cases, the fourth STC we look at (Fig.~\ref{fig:all_couplings}(d)) is a simple extension of the previous two couplings, where the spatio-spectral phase in the NF is now quadractic in frequency rather than linear (Fig.~\ref{fig:all_couplings}(d)(i)), i.e. $\hat{\phi}(x,y,\omega)=\zeta x (\omega-\omega_0)^2$. This can be understood as a transversely-varying linear temporal chirp, which in time corresponds to a transversely-varying pulse duration. This is also equivalent to the different colors having a wavefront tilt that varies quadratically with the frequency offset. In the focus this frequency-varying tilt manifests as a quadratically-varying best-focus position in the transverse dimension (Fig.~\ref{fig:all_couplings}(d)(iii)). In time at focus the pulse amplitude is quite complex, but the temporal phase no longer exhibits any chirp, because the chirps of different signs in the NF average-out at focus.

The summaries above and the quantitative visualizations in Fig.~\ref{fig:all_couplings} are on one hand relatively simple, and of low-order, but on the other hand can be quite difficult to digest in one sitting. However, understanding the difference between the spatio-spectral phases employed and the reasoning behind the relationships between time and frequency and also NF and FF is key to understanding STCs. This is true both of low-order STCs and those of arbitrary nature. All four of the cases in Fig.~\ref{fig:all_couplings} have unique effects in focus and also in time on the collimated beam, but have identical spatio-spectral amplitudes in the NF. In the NF, it is \textit{only} the spatio-spectral phases  that differentiate them. From a practical point of view this makes sense, since we often imagine STCs being induced on the collimated beam in the form of chromatic phase aberrations, but in the general case we must also be open to more complex field configurations. In the next section, we briefly discuss some physically relevant cases which involve more complex couplings in the spectral domain.

\subsection{Examples of more complex couplings}
\label{sec:concepts_complex}

\begin{figure}[htb]
	\centering
	\includegraphics[width=83mm]{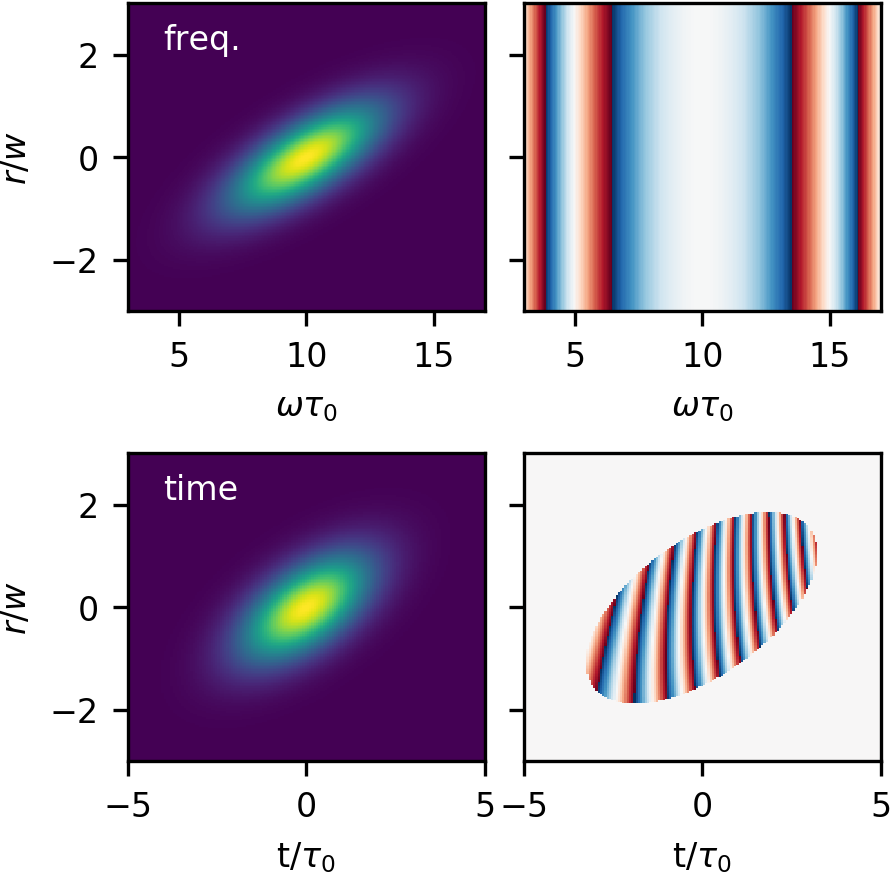}
	\caption{A transversely varying central frequency and a spatially-homogeneous linear temporal chirp (quadratic spectral phase), shown in the top row, produce pulse-front tilt in time, shown in the bottom row. However, the field is different than the 'standard' AD/PFT, despite the PFT in both cases. The color scale for phase goes from $-2\pi$ to $2\pi$.}
	\label{fig:PFT_akturk04}
\end{figure}

A simple example of pure amplitude coupling in the spectral domain and NF is the case of a beam that has been compressed by a single-pass grating compressor. In this case, the beam central frequency varies linearly with the transverse position, i.e. it has transverse chirp in the NF. If an overall temporal chirp is applied to such a beam, for instance by moving one of the gratings in the compressor, then the combination of these two effects obviously results in a tilt between the pulse-front and the wavefront~\cite{akturk04} (Fig.~\ref{fig:PFT_akturk04}). This is the same temporal intensity effect as in the 'standard' AD/PFT, yet with a field configuration that is actually different both in the NF and FF. Hence this example is very instructive from the point of view of STC metrology, since a measurement of only the spatio-temporal intensity would not provide information on the full nature of the beam.

It is important to stress as well that amplitude couplings can spontaneously arise when pulses are very broadband, even in the simple case of a freely propagating beam, due to the chromatic character of diffraction and propagation~\cite{feng98,porras02}. An example is the case of a broadband cavity operating with Kerr-lens mode-locking, where the Rayleigh range is fixed to be the same for all frequencies. This results in a beam-size that varies according to $\sqrt{\lambda}$~\cite{cundiff96}. This type of effect can become very significant when pulses approach the few-cycle limit, affecting even the Gouy phase and central frequency through a focus~\cite{porras09,hoff17}.

Amplitude couplings can also easily occur in the misalignment of non-collinear OPAs~\cite{harth18}. Beyond low-order couplings, high order couplings can have a myriad of sources, for example due to changes in laser gain medium~\cite{tamer18} or temporal gain dynamics in highly-saturated Joule-level amplifiers~\cite{jeandet19}.

\subsection{Simple or incomplete measurement techniques}
\label{sec:concepts_measurements}

Before discussing advanced techniques, it is useful to discuss some experimentally simple techniques that can determine whether certain couplings are present, although not necessarily precisely their magnitudes. This is useful since these methods generally require very little specific or expensive devices and are therefore very accessible, and also apply to many of the real-world scenarios that scientists may encounter.

The most well-known of these simple techniques is to diagnose the focus of an ultrashort laser beam with the full spectrum and compare to that with a narrow central part of the spectrum. In practice, this can be achieved using an appropriate band-pass spectral filter, placed in front of the sensor used to measure the focal spot profile. Referencing Fig.~\ref{fig:all_couplings} can already hint that for both AD/PFT and CC/PFC in the NF, the effects of the couplings should be easily visible at the focus (FF). For AD/PFT the focus will be extended in one direction, i.e. elliptical, but will be round with only a narrow part of the original spectrum. Similarly a beam with CC/PFC will be larger than the expected diffraction-limited spot size in focus, but will get closer to this expectation with only a narrow part of the spectrum. In both cases an achromatic aberration may at first be suspected, for example astigmatism causing an elliptical focus, but the different nature of the focus with a narrower spectrum can make it possible to distinguish between chromatic and achromatic aberrations.

\begin{figure}[htb]
	\centering
	\includegraphics[width=83mm]{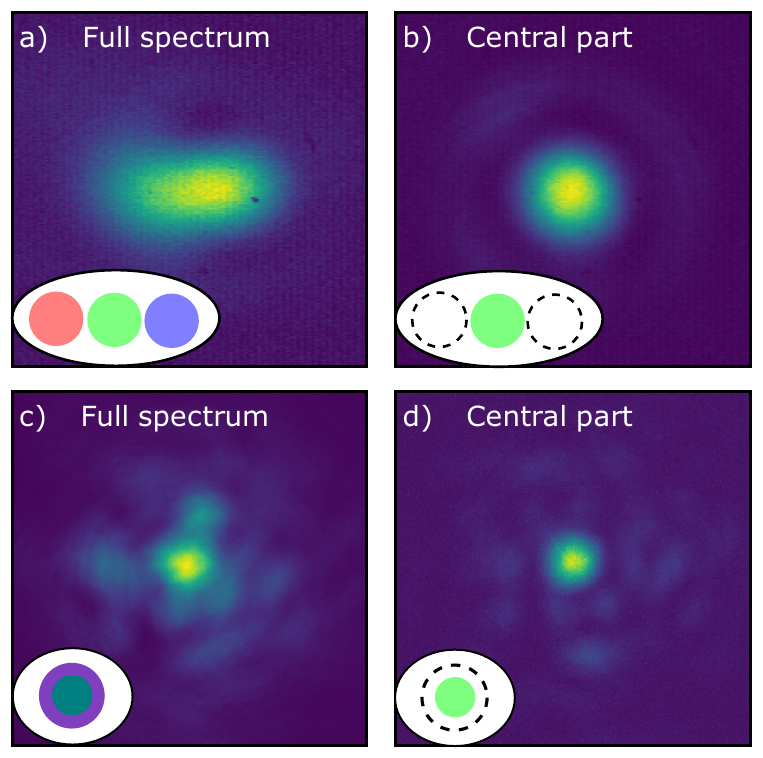}
	\caption{An example of simple diagnostic of STCs, applied to two different cases. (a) and (b) are from the same focused beam with PFT, but (b) has a band-pass filter in front of the camera, while (a) is a measurement with the full beam spectrum. (c) and (d) compare the same types of measurements, now for a beam with PFC. In these two examples, the focus with the band-pass filter added shows the high quality of the focus, but the focus with the full spectrum reveals the coupling. Both sets of data are from different 800\,nm laser systems with large spectra, which have different focusing conditions. (c) and (d) are adapted from Ref.~\cite{jolly20-1} \textcopyright The Optical Society.}
	\label{fig:STC_focus_exp}
\end{figure}

Experimental examples of this for AD/PFT and CC/PFC in the NF (with different lasers and focusing conditions) can be seen in Fig.~\ref{fig:STC_focus_exp}. Due to transverse chirp, the elliptical focus obtained with the full spectrum in Fig.~\ref{fig:STC_focus_exp}(a) is revealed to be round with only the central part of the spectrum in Fig.~\ref{fig:STC_focus_exp}(b). Due to longitudinal chirp, the large beam in Fig.~\ref{fig:STC_focus_exp}(c) is revealed to be smaller and more round with only the central part of the spectrum in Fig.~\ref{fig:STC_focus_exp}(d). In this latter case of CC/PFC the beam with the entire spectrum (Fig.~\ref{fig:STC_focus_exp}(c)) is more complex since the NF had a flat-top profile, so the frequencies not at best focus do not have a simple spatial distribution.

Although this technique does show the presence of a coupling, only after a complex convolution of the spectrum and the measured profiles with and without the band-pass filter could one expect to quantify the coupling. Still, it can be a very useful technique due to the simplicity. This is why, when the source of the coupling is known and it is a simple step to tune the value, such a method can be very practical and useful. For example, minimizing the ellipticity of a beam with the full spectrum can be an indirect measure for minimizing AD/PFT when using a prism or grating compressor, where it is known that AD/PFT is very easily induced by misalignement. However, when the situation is more complex it cannot give much information, especially when the spectrum contains many features or there are a combination of multiple STCs and/or achromatic focusing aberrations.

\begin{figure}[htb]
	\centering
	\includegraphics[width=83mm]{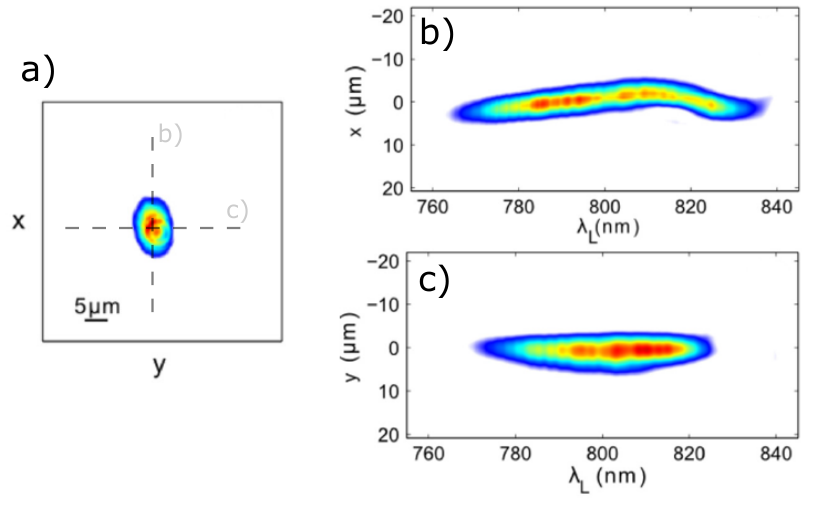}
	\caption{Measurements of the spectrally-resolved focal spot profile along two slices in the FF of a 100\,TW laser beam (a). The slice shown in (b) has a quasi-parabolic dependence of central frequency on position, where the slice in (c) shows no significant STC. Results are taken from Ref.~\cite{kahaly14} with permission.}
	\label{fig:kahaly14}
\end{figure}

A further advancement of such a measurement was undertaken on a 100\,TW laser system and produced meaningful results, which was simply using an imaging spectrometer to spectrally resolve the focal spot along both spatial axes~\cite{kahaly14}. Further extensions of this simple approach would consist in spatially scanning the beam in two dimensions with a fiber spectrometer, or scanning in one dimension with an imaging spectrometer, in order to reconstruct the full spatio-spectral amplitude. Although useful, the weakness of such measurements is that they do not provide any information on the spatio-spectral phase in the NF or FF. The measurement results of Ref.~\cite{kahaly14}, shown in Fig.~\ref{fig:kahaly14}, provide an interesting illustration of this limitation. A curved spatio-spectral amplitude was observed in the FF, qualitatively similar to the case of the last coupling of Fig.~\ref{fig:all_couplings}(d). Yet, since the spatio-spectral phase remained unknown, there was no way to experimentally verify that the measured beam distortion was actually due to a spatially-varying temporal chirp in the NF. As a consequence, unambiguously identifying the nature and physical origin of this distortion in the laser system turned out to be impossible. 

There are many other examples of such 'incomplete' measurements, of varying complexity, which produce results that are not complete representations of the pulse electric field. These include interferometric measurement of radial group-delay~\cite{bor89-2,netz00}, extensions of single-shot autocorrelation to measure pulse-front tilt~\cite{pretzler00,sacks01,akturk03,figueira19} or pulse-front curvature~\cite{wu16}, multiple-slit spatio-temporal interferometry~\cite{li18-2,li19}, There are more advanced diffractive methods that can do similar analysis, using a structured diffraction grating, referred to as "chromatic diversity"~\cite{bahk18}, or a measurement of angular chirp simultaneously in both spatial dimensions~\cite{osvay05,borzsonyi13}. There are also methods that are interested in only the temporal intensity profile (including the absolute intensity magnitude), for example the Temporally-Resolved Intensity Contouring (TRIC) technique~\cite{haffa19}. Although of interest, such methods will not be discussed further in this tutorial.

\section{Spatially-resolved spectral measurement techniques}
\label{sec:spatial}

We first address techniques that we deem are spatially-resolved spectral/temporal measurements, i.e. measurements that resolve the spectrum and spectral phase, extended to one or more spatial dimensions in order to resolve STCs. It is very important to emphasize that, maybe counter-intuitively, simply adding spatial resolution without caution to one of the usual techniques for purely temporal measurements, for instance by scanning the measurement device over space, actually does not provide a full spatio-temporal characterization of a laser beam. This is due to the fact that these techniques generally only measure the components of the spectral phase that affect the pulse duration and shape, but are blind to those that are constant or linear in frequency---which respectively correspond to the Carrier-Envelope relative Phase, and to the pulse arrival time. Therefore a device scanned across space would be able to detect the spatially varying envelope (due either to a change in nonlinear components of the spectral phase or to a change in the spectral width), but not something as simple as AD/PFT or CC/PFC, where the pulse shape does not vary at all spatially.

\begin{figure}[htb]
	\centering
	\includegraphics[width=83mm]{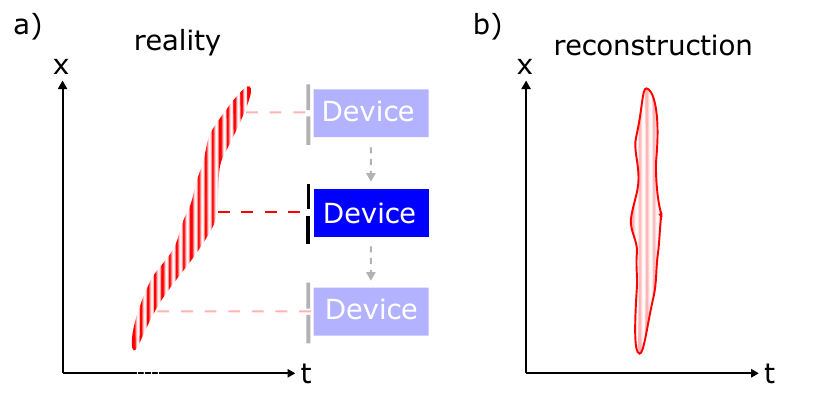}
	\caption{Sketch explaining the necessity of spatial-temporal characterization. Since the temporal characterization device ("Device") in this case is blind to spatial variations of the carrier phase and absolute arrival time of the true pulse in (a), the pulse-front tilt cannot be resolved in the reconstructed pulse shown in (b).}
	\label{fig:STC_need}
\end{figure}

This important idea is illustrated via a sketch in Figure~\ref{fig:STC_need}, where the rastering of the device can resolve the more nuanced fluctuations in pulse length, but not the pulse-front tilt. Of course this may already be a useful amount of information, for example in pulse broadening in a plasma~\cite{zair07,beaurepaire16}, but it is not a complete measurement. Furthermore, in practice the rastering process is itself limited due to the large number of measurements necessary to have a high resolution, especially if the measurement is performed on both transverse dimensions. 

We outline three different types of measurements that rely on spatially-resolving various methods of pulse characterization. Some of them make it possible to avoid the previous issue, while this can be a very tricky problem for others. The techniques are all based on forms of interferometry, so in every case the unknown beam needs to interfere either with a known reference, or with itself (so-called self-referenced interferometric techniques). One of the key challenges of this category of techniques is precisely to find ways to generate an appropriate reference beam.

These techniques can be differentiated mainly by the type of reference used, and the method for resolving the measurement spatially and spectrally. These include: self-referenced techniques such as SPIDER or SRSI, resolved on an imaging spectrometer (section~\ref{sec:SPIDER}, 'established techniques extended to spatial dimensions'); spectral interferometry raster-scanned over the spatial extent of a beam, which can be either externally-referenced or be referenced to a single point on the unknown beam (section~\ref{sec:SEA-TADPOLE}, 'spatially-resolved spectral interferometry); and self-referenced Fourier-transform interferometry, where a spatially-extended reference is made from some central portion of the unknown beam and spectral resolution is obtained via Fourier-Transform spectroscopy (section~\ref{sec:TERMITES}, 'spatially-resolved Fourier-transform interferometry'). 

\subsection{Established techniques extended to spatial dimensions}
\label{sec:SPIDER}

\begin{figure*}[htb]
	\centering
	\includegraphics[width=171mm]{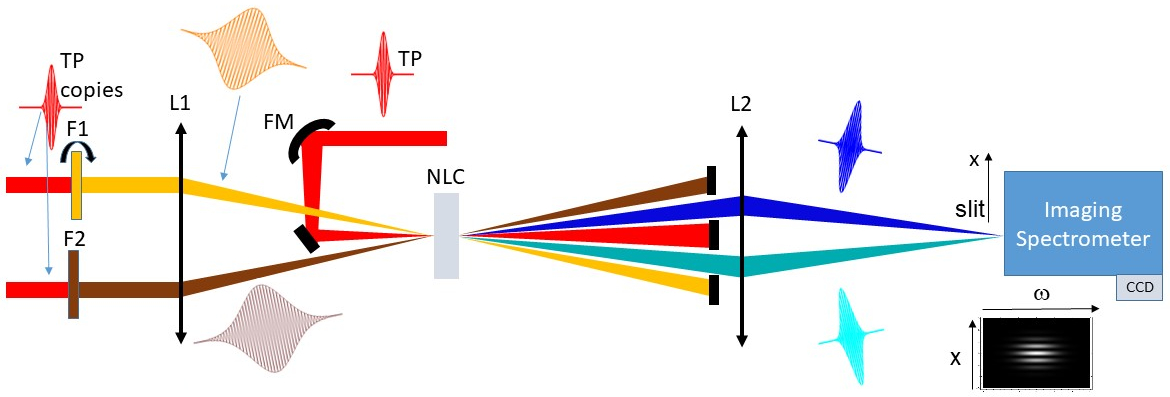}
	\caption{SEA-F-SPIDER setup. A test pulse (TP) is split into three copies, two of which create narrowband ancillae using filters (F1 and F2). The TP and ancillae are focused in a nonlinear crystal (NLC) with curved mirror FM and lens L1 respectively, with relative angles, producing two beams via sum-frequency generation (SFG) that are shifted in frequency and at distinct outgoing angles. The directly transmitted TP and ancillae are blocked (along with the potential second-harmonic of each) and the signal beams are re-focused (with L2) on to an imaging spectrometer having the slit in the plane of the page (along $x$). F1 is rotated in order to produce the zero-shear measurement, which is for calibrating the angle of arrival on the spectrometer (represented by $k_x$). The version shown is measuring the STCs of the TP in the far-field, but if there were no focusing optics (lenses $L_1$ and $L_2$) then it could in principle characterize directly the near-field.} 
	\label{fig:SEA-SPIDER}
\end{figure*}

In this section we will discuss established techniques extended to spatial dimensions, including mainly the SPIDER and SRSI techniques. In each case the technique is expanded to one spatial dimension only, which already limits the information it provides. Regarding which STCs are accessible by this class of technique, there is some ambiguity in the literature, but we will provide our perspective here.

SPIDER is a self-referenced interferometric technique for spectral/temporal measurements, where the reference beam consists of a spectrally-sheared version of the test pulse (TP) to be characterized.  Implemented with a 1D spectrometer, this provides the local spectral amplitude and phase of the field in one shot~\cite{iaconis98}. An obvious extension of this technique consists in rather using a 2D imaging spectrometer to obtain spatial resolution along one spatial axis. Historically, this has been one of the first approaches implemented for STC measurements ~\cite{gallmann01}, and this is why we discuss it first.  We will however show that all spatially-resolved versions of SPIDER are affected by the limitation illustrated in Fig.~\ref{fig:STC_need}, and explain how only a more sophisticated measurement scheme can circumvent this limitation ~\cite{dorrer02-3,dorrer02-1,dorrer02-2}. 

In order to understand the subtle issues involved in the generalization of SPIDER to STC measurements, it is useful to provide a more detailed description of the technique. To this end, we will focus on the particular implementation called SEA-F-SPIDER~\cite{witting09-1}, sketched in Fig.~\ref{fig:SEA-SPIDER}, both because this is the latest and most advanced version, and because it has been used for the spatio-temporal characterization of few-cycle near-infrared and mid-infrared sources ~\cite{witting12,austin16,witting16,witting18}. For a review of the historical development of SPIDER and its different versions, and of the numerous practical advantages of SEA-F-SPIDER we refer the reader to Ref.~\cite{Monmayrant_2010}. 

In any SPIDER device, the key operation is to create two replicas of the TP, which are sheared in frequency by a fraction of the TP's spectral width. This is typically achieved by performing sum frequency generation of this test pulse in a second-order nonlinear crystal, with two quasi-monochromatic waves of slightly different frequencies. In SEA-F-SPIDER, these two waves are generated by producing two ancillary beams, obtained by passing two samples of the TP through  separate narrowband spectral filters, placed at slightly different angles. For STC measurements, these ancillae can also be spatially filtered to avoid spatio-temporal distortions of the TP in the frequency conversion process ~\cite{witting09-1,wyatt11}. 

The technique then consists in comparing these two replicas, which is achieved by spectrally-resolved interferometry. As sketched in Fig.~\ref{fig:SEA-SPIDER}, in SEA-SPIDER, the two replicas are recombined at an angle on the entrance slit on an imaging spectomerer, creating spatial fringes which encode the difference in spatio-spectral phase between the two beams. The mathematical expression of the interferogram measured in this scheme is: 
\begin{align}
\begin{split}
&S\left(x,\omega\right)=|\hat{E}\left(x,\omega\right)|^2+|\hat{E}\left(x,\omega-\Omega\right)|^2 \\
&+2|\hat{E}\left(x,\omega\right)\hat{E}\left(x,\omega-\Omega\right)| \\
&\times\cos\left[\hat{\phi}\left(x,\omega\right)-\hat{\phi}\left(x,\omega-\Omega\right)-k_x x\right] \label{eq:SEA-SPIDER},
\end{split}
\end{align}
\noindent with $\Omega$ the induced spectral shear, and $k_x$ the transverse wave vector difference between the two beams. A straightforward processing, based on Fourier transformations and filtering and commonly used in interferometry, makes it possible to get the phase $\Delta \varphi(x,\omega)=\hat{\phi}\left(x,\omega\right)-\hat{\phi}\left(x,\omega-\Omega\right)-k_x x$ from $S\left(x,\omega\right)$, provided $k_x$ is large enough. 

The next step, common to all implementations of SPIDER and called the calibration procedure, is to eliminate the phase term responsible for the interference fringes, which in the case of SEA-SPIDER is $-k_x x$. This phase term can be determined and then subtracted by performing a measurement of $S\left(x,\omega\right)$ without the spectral shear~\cite{kosik05}, leading to $\Delta \varphi(x,\omega)=-k_x x$. In SEA-F-SPIDER, this is achieved by simply rotating one of the spectral filters, so that both ancillae have the same frequency and no shear is induced between the two interfering replicas of the TP. 

We emphasize that this calibration step is crucial for SPIDER, and has to be performed with great care to make sure that no extra phase term is introduced in the procedure. This can become particularly tricky for STC measurements: in Ref.~\cite{wyatt09}, it was found that the SFG process of the TP with the spectrally-sheared ancillae in a non-collinear geometry introduces an extra phase term compared to the calibration configuration. This extra phase term corresponds to a spurious angular dispersion, and needs to be corrected for to get meaningful results.

After this calibration step, and assuming no extra phase term has been introduced, one gets $\hat{\phi}\left(x,\omega\right)-\hat{\phi}\left(x,\omega-\Omega\right)\approx \Omega \;\partial \hat{\phi}\left(x,\omega\right)/\partial \omega$. The spatio-spectral phase $\hat{\phi}\left(x,\omega\right)$ can then be obtained by an integration with respect to frequency, obviously up to an unknown frequency-independent phase term. In other words, this type of technique provides no information on the frequency-average wavefront of the beam. As we now explain, this unknown phase term prevents the determination of certain types of STC, such as pulse front distortions \textsl{in the measurement plane}.

To demonstrate this point, we consider the simple case of a beam with AD/PFT in the measurement plane. As explained in section~\ref{sec:concepts_manifestations} and demonstrated in ~\ref{sec:appendixA}, such a beam is described in the spatio-spectral domain by a phase $\hat{\phi}\left(x,\omega\right)=\gamma (\omega-\omega_0) x$, with $\omega_0$ the central frequency of the beam. A SPIDER measurement will then provide $\partial \hat{\phi}\left(x,\omega\right)/\partial \omega= \gamma x$, whatever its specific implementation and assuming a perfect calibration procedure. It is then tempting to conclude that information about this coupling is indeed obtained, but this conclusion omits an essential point. 

As explained in ~\ref{sec:appendixA}, a perfect STC-free beam that propagates at an angle (here with respect to optical axis of the SPIDER device) is described by a phase $\hat{\phi}\left(x,\omega\right)=\gamma \omega x$, which only differs from the phase of the beam with AD/PFT by a term independent of frequency. For such a beam, a SPIDER device also measures $\partial \hat{\phi}\left(x,\omega\right)/\partial \omega= \gamma x$. This shows that SPIDER cannot distinguish between a beam with AD/PFT and a simple tilted beam, due to the fact that it retrieves the phase up to an unknown frequency-independent term. Following the same reasoning, it is clear that it cannot either distinguish a beam with PFC from a converging or diverging beam. This is precisely the type of limitation illustrated in Fig.~\ref{fig:STC_need}.

Although this is a conceptual limitation of this technique, it is worth stressing that in most practical cases, this is probably not a significant shortcoming. Indeed, in SPIDER it is generally the laser field in the plane of the SFG crystal that is characterized. This corresponds to the far-field of the laser beam, while couplings such as PFT, PFC or other pulse front distortions typically occur in the NF and the nature of STCs change from NF to FF as explained in section~\ref{sec:concepts_manifestations}. Techniques like SEA-F-SPIDER can therefore still provide very useful information on STC at focus. The measurement at focus of the peculiar PFT resulting from the combination of transverse spatial chirp and temporal chirp has for instance indeed been demonstrated~\cite{witting16}.

Overcoming this general limitation of the technique is possible, however, by combining a spatially-resolved SPIDER with a spectrally-resolved spatial-shearing interferometer~\cite{dorrer02-3}, in order to be able to measure not only $\partial\hat{\phi}/\partial\omega$ but also $\partial\hat{\phi}/\partial{x}$~\cite{dorrer02-1,dorrer02-2}. This scheme is sometimes called 2D-SPIDER. From these two measurements it is possible to reconstruct the entire spatio-spectral phase along one spatial dimension. Considering again the comparison of a beam with AD/PFT and a tilted beam, these two cases can now be distinguished, since $\partial\hat{\phi}/\partial{x}=\gamma (\omega-\omega_0)$ in the first case, while $\partial\hat{\phi}/\partial{x}=\gamma \omega$ in the second. The technique is potentially single shot, but requires a rather complex optical set-up and a precise and very careful calibration, and has therefore not been in common use so far.

SRSI~\cite{oksenhendler10,moulet10,trisorio12} is another technique which is commonly used to determine the electric field of a pulse in time. In this technique, a nonlinear $\chi^{(3)}$ effect (cross-polarized wave generation, XPW~\cite{minkovski04}) generates a temporally-filtered replica of the input TP, which is then used as a reference pulse for interferometric measurements. The key idea is that this reference pulse, which is spectrally broader than the TP, can be considered to have a nearly flat spectral phase. Imperfections of this reference pulse can be taken into account through the use of an iterative algorithm~\cite{oksenhendler10}. 

In order to be able to obtain spatially-resolved measurement and potentially STCs, a tilt between the TP and the reference is implemented (SRSI-ETE)~\cite{oksenhendler17}, and the interferogram is measured with an imaging spectrometer. The interferogram obtained in this case at the sampled position $y_0$, with $k_x$ and $\tau$ representing the angle and temporal delay respectively, is given by:
\begin{align}
\begin{split}
&S\left(x,\omega\right)=|\hat{E}\left(x,\omega\right)|^2+|\hat{E}_{\textrm{XPW}}\left(x,\omega\right)|^2 \\
&+2|\hat{E}\left(x,\omega\right)\hat{E}_{\textrm{XPW}}\left(x,\omega\right)| \\
&\times\cos\left[\hat{\phi}\left(x,\omega\right)-\hat{\phi}_{\textrm{XPW}}\left(x,\omega\right)-k_x x-\omega\tau\right] .
\end{split}
\end{align}
\noindent If the spectral phase of the reference is properly filtered by the XPW process, and with a proper calibration of $k_x x + \omega\tau$, this interferogram can be used to retrieve $\hat{\phi}\left(x,\omega\right)$. Yet, this technique should suffer from the same ambiguities as the SPIDER technique, again related to the issue emphasized in Fig.~\ref{fig:STC_need}: the XPW process used to generate the reference does not filter the constant and linear terms of the spectral phase, corresponding in space-time to wave front and pulse front. As a result, the SRSI-ETE technique should not be able to resolve pulse-front distortions such as AD/PFT and CC/PFC. We note however that Ref.~\cite{oksenhendler17} claimed a measurement of the former, which might have been possible in this specific implementation because the XPW process was carried out in the FF, and the SRSI-ETE measurement in the NF. The form of PFT due to the combination of spatial and temporal chirps would be resolved, since it is due to second-order spectral phase that would be filtered on the reference by the XPW process. 

There other examples of devices expanding upon the FROG technique, ImXFROG~\cite{eilenberger13} and HcFROG~\cite{mehta14}. However they have had very limited use and therefore fall outside of the scope of this tutorial.

All the techniques described here have their own merits, but also tend to be experimentally complicated and require very precise calibration. Indeed, for various ambiguities it is not completely clear experimentally what the limits of the techniques are, and because they are self-referenced it is difficult to set a threshold for when the calibration has been done properly. So one should take great consideration when choosing if a technique in this section is suitable for one's application.

\subsection{Spatially-resolved spectral interferometry}
\label{sec:SEA-TADPOLE}

\begin{figure*}[htb]
	\centering
	\includegraphics[width=171mm]{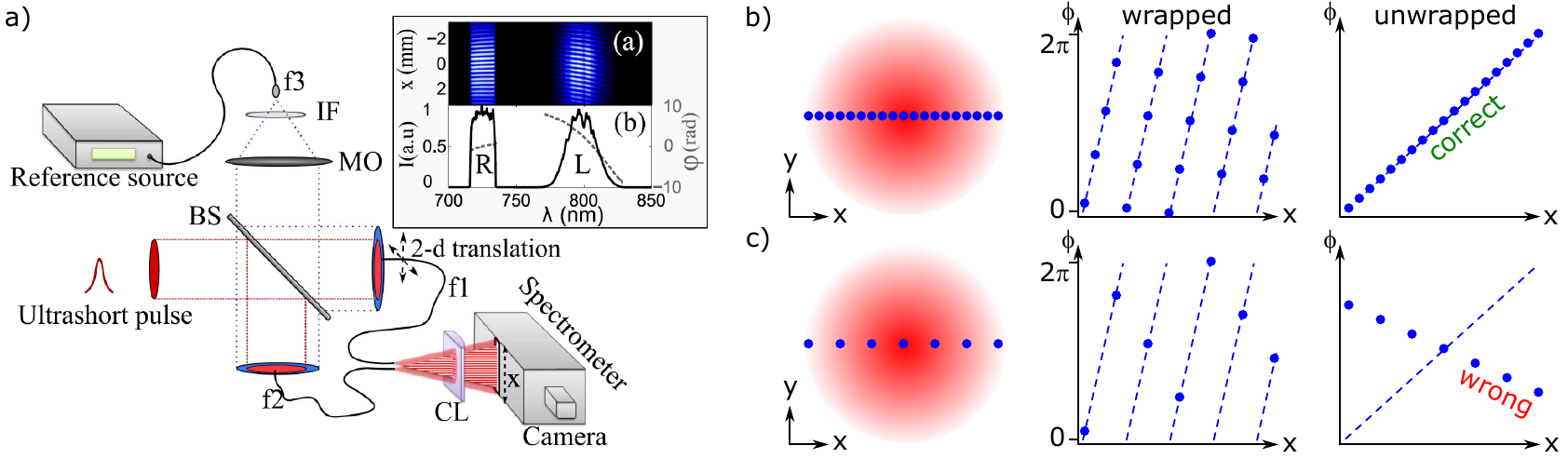}
	\caption{(a) RED-SEA-TADPOLE, an evolution of the SEA-TADPOLE that involved spatially-rastered spectral interferometry with a spectrally distinct reference to account for imperfections in the 2-D stage and the fibers. When the device is rastered across the tranverse dimension with high enough resolution as on the left in (b) the unwrapped phase will correspond to the true case (dashed line), shown on the right. When the rastering is done with much lower resolution as on the left in (c), the wrapped phase will be ambiguous, shown in the center. The risk is that the unwrapped phase is completely wrong,as shown on the right. For both (b) and (c) the true phase (dashed lines) is simply linear, which could be from a simple tilt between the scanned plane and the plane perpendicular to the laser propagation. The image in (a) is reprinted with permission from Ref.~\cite{gallet14-2} \textcopyright The Optical Society.}
	\label{fig:SEA-TADPOLE}
\end{figure*}

Spectral interferometry between an unknown pulse and a \textsl{known reference pulse} can provide the full spectral amplitude and phase information of the unknown pulse, and is a component of many standard temporal pulse characterization devices that have matured significantly to this point in time. We will outline devices that involve spatially-resolving spectral interferometry measurements in order to reconstruct the 3-D electric field, which are referred to as SEA-TADPOLE, STARFISH, and RED-SEA-TADPOLE. These are very different from the SEA-SPIDER and SRSI-ETE methods, in that they have no spectral or spatial shear, and therefore require an additional spectral phase measurement to reconstruct the complete spatio-temporal electric field, and spatial resolution is obtained by scanning the beam with an optical fiber.

The original TADPOLE technique was developed to temporally characterize very weak pulses using a combination of a FROG measurement for a reference beam, and a single spectral interferogram between the reference and an unknown pulse~\cite{fittinghoff96}. SEA-TADPOLE was then developed as a variant where the unknown and reference pulses are collected by monomode fibers, and then compared by spatio-spectral interferometry~\cite{meshulach97} between the two separate fiber outputs~\cite{bowlan06}. This was then naturally extended to spatio-temporal metrology by scanning one of the two fibers' input across the spatial extent of the unknown beam, to measure its spatially-resolved spectral amplitude and phase~\cite{bowlan07} through comparison with the fixed reference beam. The first measurements with this technique could achieve quite high spectral and spatial resolutions~\cite{bowlan08}. The STARFISH technique~\cite{alonso10,alonso12,alonso13} is essentially a simple variant of the same technique, which simplifies the experimental implementation by rather using standard fiber spectrometers and a fiber coupler, and relying on pure spectral interferometry, thus avoiding the need for a 2D spectrometer. In most practical cases, the reference beam is obtained from a part of the unknown beam itself, which is then separately characterized in time. We emphasize that this reference must cover the full spectral extent of the unknown beam, which can be an issue for the characterization of beams with strong inhomogeneities in spectral amplitude.

A critical experimental issue is related to phase fluctuations in the monomode fibers, due e.g. to vibrations or temperature fluctuations. These are particularly difficult to avoid in these measurement schemes as at least one of the fibers needs to be scanned spatially---which necessarily implies a minimum amount of deformation. The main effect will generally be a randomly-varying overall phase term induced on the spectral phase of the test pulse as the beam is scanned. This implies that the wave front of the beam cannot be retrieved, and hence than STC such as PFT, PFC or any other pulse front distortions cannot be measured. A solution to this issue has been proposed and demonstrated in Ref.~\cite{bowlan12}, where SEA-TADPOLE measurements were done in multiple $z$ planes, and a standard phase retrieval algorithm (see section~\ref{sec:phase_retrieval}) was then applied to the measurement results to retrieve the wavefront. This however requires scanning the beam transversely (in principle in 2D) with the fiber in \textsl{multiple planes}.

The SEA-TADPOLE results cited here were generally applied on a focused beam or a beam near the focus, which is why the phase-retrieval technique proposed in Ref.~\cite{bowlan12} was possible. However, if the beam to be measured has a pointing jitter such that fluctuations in the focus position become a significant fraction of the focal spot size, sampling this focus by scanning an optical fiber obviously becomes meaningless. Such a situation typically occurs on high-power ultrashort lasers. It is then necessary to rather scan the beam in the near-field, where the spatial extent is significantly larger. With a larger scanning range, the issues of phase fluctuations in the fiber, and of the stability and accuracy of the rastering stage and of the whole interferometer become much more critical.

An evolution to the SEA-TADPOLE device, named RED-SEA-TADPOLE~\cite{gallet14-2}, was developed to solve these concerns and thus be able to apply this type of techniques to the collimated beams of e.g. high-power ultrashort lasers. The RED-SEA-TADPOLE device, shown in Fig.~\ref{fig:SEA-TADPOLE}(a), uses a second reference beam which needs to fulfill stringent conditions (detailed in Ref.~\cite{gallet14-2}), the main ones being that it must be in a spectral range different from that of the test beam, and that it has to be free of STCs. This reference beam is collected by the two fibers together with the unknown beam. The purpose of this spectrally distinct reference is to be able to independently measure imperfections in the spatial scanning and fluctuations in the fibers and to subtract them from the final measurement of the unknown beam. Essentially, since the spectrally distinct reference is homogeneous and free of STCs, any distortion of this beam retrieved by the measurement must be due to the stage wobbling or the fiber fluctuations. The production of a suitable spatially-extended reference beam with sufficient photon flux is however far from trivial. This was done in Ref.~\cite{gallet14-2} with an expanded photonic-crystal fiber-based supercontinuum source and a band-pass filter outside of the band of the unknown laser pulse spectrum.

So with the RED-SEA-TADPOLE device a large collimated beam can now be measured with all scanning imperfections accounted for. In Ref.~\cite{gallet14-2} this was successfully utilized to measure both AD/PFT and CC/PFC due to a detuned grating compressor and a chromatic lens respectively. However, there remains one important issue that is relevant to SEA-TADPOLE, STARFISH, and RED-SEA-TADPOLE (despite the improvements in the latter). That is, due to ambiguities in phase-unwrapping, the spatial scanning resolution necessary to truly resolve an unknown pulse in the nearfield are very stringent.

This phase unwrapping ambiguity is shown in Fig.~\ref{fig:SEA-TADPOLE}(b-c). If the measured pulse is simply tilted with respect to the plane of the rastering, then the true phase across one spatial cut will be linear with a slope depending on the tilt (shown in the dashed lines). This is relevant because in practice it is extremely difficult to align the plane of a 2D stage exactly with the plane perpendicular with the laser propagation. The raw measured phase, which is necessarily wrapped, will require unwrapping to resolve this tilt. When the beam is scanned with high resolution as in Fig.~\ref{fig:SEA-TADPOLE}(b), there is no ambiguity in the phase unwrapping, so the result is correct. However, when the scanning resolution is much lower as in Fig.~\ref{fig:SEA-TADPOLE}(c), the wrapped phase is no longer unambiguous, because it is not known a priori how many $2\pi$ phase jumps occur between each measurement point. When the phase is unwrapped there is a high likelyhood of producing an incorrect unwrapped phase. This is a general issue in phase unwrapping, but in the context of RED-SEA-TADPOLE combined with a large beam in the NF, this implies that rastering in both tranverse dimension requires millions of sample points. This issue means that for practical purposed SEA-TADPOLE on a collimated beam is often just too inconvenient to implement. However, for beams in-focus it is still a reasonable solution, because in such a configuration having a very high spatial sampling is practically achievable~\cite{bowlan08}. We however note that in all measurements performed with these techniques so far, the beam was in practice sampled along one spatial direction only, due to the burden of finely scanning the fiber tip in 2D. 

As a last point, it is theoretically possible to create a multiplexed version of SEA-TADPOLE for measurements in the near field, to reduce or eliminate the need for scanning in 2D, referred to as MUFFIN in Ref.~\cite{gallet14-1}. But in this case the same difficulties regarding the phase-jumps in space exist, and the experimental complexity of adding fibers and the lack of high spatial resolution in the final data make it not very attractive in the end, and in fact it was never implemented for the characterization of STCs.

\subsection{Spatially-resolved Fourier-transform spectroscopy}
\label{sec:TERMITES}

A major drawback of the previous techniques is that the unknown beam needs to be scanned spatially in 2D to measure the full 3D spatio-temporal or spatio-spectral field. It would obviously be more straightforward to directly resolve the two transverse spatial coordinates  of the beam on a camera. But one then needs to find a way to resolve the third coordinate, i.e. to get spectral resolution. We now explain how this can be achieved by exploiting Fourier-transform spectroscopy (FTS), leading to different techniques where only one physical parameter---a temporal delay- now needs to be scanned, described in this section and later in section~\ref{sec:phase_retrieval}. 

FTS is a powerful method to resolve the spectrum of an unknown beam by measuring the evolving signal on a photodiode as the unknown beam temporally interferes with a delayed copy of itself. The resulting interferometic trace, composed of the signal measured at all of the scanned delays $\tau$, contains the first-order autocorrelation function of the field, and can be Fourier-transformed to frequency in order to directly produce the spectral intensity $\hat{I}(\omega)$, when selecting only the spectral information at the positive frequency peak. More explicitly:

\begin{equation}
\hat{I}(\omega)=\left\lbrace \mathcal{F}_{\tau\rightarrow\omega}\left[\int \left|E(t)+E(t-\tau)\right|^2 dt\right]\right\rbrace_{+\omega} ,\label{eq:FTS}
\end{equation}

\noindent which is essentially the Wiener-Khinchin theorem~\cite{doi:10.1002/0471213748.ch11}. Since the beams interfering are copies of each other, the resulting beam has a zero spectral phase regardless of the phase on the input beam. This has the benefit that the spectral phase of the beam to be measured does not matter, but of course the downside is that FTS cannot resolve the spectral phase. Since FTS is a scanning method there are significant implications of shot-to-shot fluctuations and noise on the resulting spectrum~\cite{dorrer00}. This makes FTS generally less preferred to simple fiber-coupled grating spectrometers (when available) to measure the 1-D spectrum, but in the case of full spatio-temporal characterization it has found a new application.

FTS can indeed be spatially-resolved quite easily, just by resolving the interfering beams on a camera. When the interfering beams are exact copies of each other, this provides a straightforward way to measure the spatio-spectral intensity. However, there is no phase information just as in the 1-D case. A self-referenced version of spatially-resolved FTS that can resolve the spatio-spectral phase was developed simultaneously and independently in ~\cite{miranda14} and ~\cite{gallet-thesis}, and later named TERMITES~\cite{pariente16}, which will be the focus of the rest of this section. TERMITES has been used successfully with varying specific experimental layouts for measurements on a Terawatt laser~\cite{pariente16}, a Petawatt laser~\cite{jeandet19}, ultrafast optical vortices~\cite{miranda17}, and a femtosecond OPCPA used for high repetition-rate high-order harmonic generation~\cite{harth18}. In each case it could resolve both standard STCs and more complex couplings in both amplitude and phase.

The TERMITES technique involves interfering the unknown beam with a spatial portion of itself, which has been expanded to overlap across the whole spatial extent of the beam. The key idea is that this expanded portion of the original beam can be used as a reference, because it comes from a small enough portion of the original beam to be free from STCs. Since the reference is then not strictly an exact copy of the unknown beam, the spatially-resolved FTS will have both amplitude and phase information, as will be explained more precisely below. Comparing TERMITES to SEA-TADPOLE, the reference in TERMITES has been expanded spatially so that on a single 2D image there is interference at all points, eliminating the need to scan in 2D that was present in SEA-TADPOLE.

\begin{figure*}[htb]
	\centering
	\includegraphics[width=171mm]{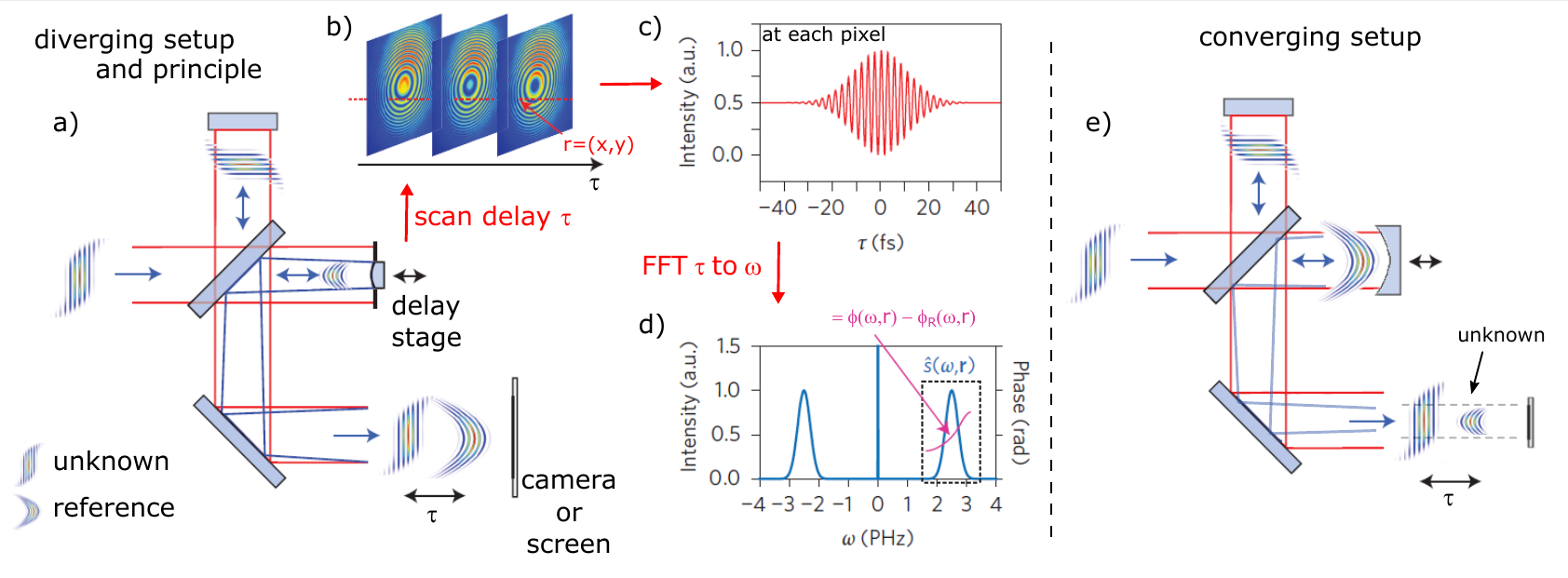}
	\caption{The TERMITES technique. Using spatially-resolved FTS and a diverging self-reference (a), the full spatio-spectral amplitude and phase can be reconstructed. The resulting interferograms (a--b) are Fourier transformed with respect to delay, to find the cross-spectral density $\hat{s}$ at each single spatial pixel, as shown in (d) at one pixel. A final iterative step produces the unknown beams spectrum and phase. A setup with a collimated reference and converging beam is also possible as in (e), requiring a smaller image size. Panels (a--d) are adapted from Ref.~\cite{pariente16} \textcopyright Nature, and panel(e) is using the same visualization style.}
	\label{fig:TERMITES_1}
\end{figure*}

An example of experimental implementation of TERMITES using a modified Michelson interferometer with one curved end-mirror is shown in Fig.~\ref{fig:TERMITES_1}(a). The overlapping beam and reference are either viewed directly on a camera chip, or put on to a scattering screen which is then imaged using a viewing objective and camera. The FTS is performed by stepping through the delay $\tau$ with steps small enough to beat the Shannon limit. Therefore significantly sub-cycle accuracy on the delay stage is essential: for characterizing 800\,nm wavelength beams we have employed steps of 150\,nm using a piezo-electric stage with fluctuations on the 10\,nm level~\cite{pariente16}.

In the end a TERMITES measurement produces a 3-D matrix of interferometric signals in $x$, $y$, and $\tau$. This data represents the spatially-resolved interferences of the unknown beam and the reference, with the addition of a large curvature value due to the fact that the reference is diverging. The first analysis steps involve Fourier-transforming from $\tau$ to $\omega$, selecting only relevant positive frequencies, and subtracting the known curvature. These steps are shown in Fig.~\ref{fig:TERMITES_1}(b-d). After these steps there remains the "cross-spectral density" $\hat{s}(x,y,\omega)$, which corresponds to the product of the complex spectral amplitude of the unknown beam, and the conjugate of the complex spectral amplitude of the reference beam. Algorithmically this corresponds to

\begin{align}
\label{termites-equation}
\hat{s}(x,y,\omega)&=\hat{E}(x,y,\omega)\hat{E}_R^*(x,y,\omega) \\
&=A(x,y,\omega)A_R(x,y,\omega)e^{\hat{\phi}(x,y,\omega)-\hat{\phi}_R(x,y,\omega)} ,
\end{align}
\noindent where the curvature of the reference beam has already been removed. At each point of the beam, one can thus get the difference in spectral phase $\hat{\phi}(x,y,\omega)-\hat{\phi}_R(x,y,\omega)$ between the unknown and reference beams. This is why we include TERMITES in the category of spatially-resolved spectral measurement techniques.

The final step of the data processing leading to the reconstruction of the unknown beam depends on the way TERMITES is implemented. In the first version utilized in Lund~\cite{miranda14}, the reference came from a very small portion of the original beam, which was moreover spatially filtered before interfering with the unknown beam. The reference beam can then be considered as originating from a point source, such that it can be assumed to be free from STC, i.e. $\hat{\phi}_R(x,y,\omega)=\hat{\phi}_R^0(\omega)$ (and similarly for the spatio-spectral amplitude). Since TERMITES is self-referenced, it is blind to any spatially homogeneous spectral phase of the unknown beam. Therefore, to reconstruct the field in the spatio-temporal domain, a single temporal measurement is still necessary, either on the reference beam or at any single point of the unknown beam (see section~\ref{sec:phase-stitching}, phase stitching). Note however that even without this final measurement, all pure spatio-spectral effects are resolved.

In the implementation presented in Fig.~\ref{fig:TERMITES_1}(a), the reference rather comes from some finite central portion of the unknown beam, due to the practical constraints imposed by the application to a laser beam of large diameter.  For instance, in the version used in Ref.~\cite{pariente16}, the reference came from the central half of the beam. This scheme is then similar to radial-shearing interferometry~\cite{doi:10.1002/9780470135976.ch5}, and the reference may itself still contain STCs. Based on the fact that it originates from a sub-pupil of the unknown beam, an iterative algorithm can be applied to the spectral amplitude and phase provided by Eq.~(\ref{termites-equation}), to eliminate the contribution of the reference and reconstruct the complex spatio-spectral field of the unknown beam. As before, a temporal measurement at a single point of the beam is still required to determine the field in space-time.

There are strict requirements on the camera properties in a TERMITES device. Due to the varying angle between the reference and unknown beam, the interference fringes increase in spatial frequency towards the outer part of the beam (see example images in Fig.~\ref{fig:TERMITES_1}(b)). The input beam diameter of the collimated beam $D$, the focal length $f$ of the convex mirror for the reference, and the fraction $\beta$ of the unknown beam diameter used to produce the reference fix the pixel size required to resolve the fringes at the outer part of the beam. If we say at the edge of the beam $p$ pixels are required for each fringe, then the linear number of pixels needed across the beam is $N=pD^2\beta/2\lambda_0|f|=pD^2(1-\beta)/2\lambda_0 L$, which leads to a total image size of at least $N^2$ pixels (see supplementary material of Ref.~\cite{pariente16}, and note the geometric constraint of $\beta=|f|/(|f|+L)$, where L is related to the total path length of the device). This can lead to requirements of tens of megapixels for beam diameters of a few centimeters, having a big impact on the data size of the final measurement (routinely many Gigabytes for a single measurement).

A nuanced distinction regarding the required pixels, which is not discussed in previous work, is that the true constraint is on the size of the pixels such that the camera signal is not averaging over a significant portion of a spatial interference fringe. If a camera chip was decimated so that fewer total pixels were chosen, but the pixel size was still small enough, then the proper signal would still be measured (although with lower resolution), now with a reduced data transfer time and smaller final data size. Additionally, the pixel size at the center of the beam could theoretically be much larger, since the constraint on pixel size is only an important constraint near the edge of the beam. But in reality both of these cases would require more advanced analysis or expensive hardware, and have not been presently investigated.

There is also a geometry of TERMITES where rather than the unknown beam staying collimated and the reference being diverging, the curved mirror can have a positive focal length and is used to reflect the unknown beam, such that the reference now remains collimated while the unknown beam converges. In this case the reference is still larger than the unknown beam on the camera, but the overall size of the relevant image is smaller. This is shown in Figure~\ref{fig:TERMITES_1}(e). The advantage of such a setup is that the size of the camera chip can be much smaller. If $\beta$ is still defined as the ratio of the unknown beam size to the reference beam size on the camera, a simple calculation shows that the requirement for the linear number of pixels across the beam in this configuration is $N=pD^2\beta/2\lambda_0|f|=pD^2\beta(1-\beta)/2\lambda_0 L$. So essentially, in this converging setup with the same device size determined by $L$, the number of pixels across the beam can be reduced by a factor of $\beta$ compared to the diverging setup, resulting in a total image with $\beta^2$ fewer total pixels. Due to the fixed geometrical relationship between $\beta$, $f$, and $L$ the focal length of the converging mirror must be longer to have a device with the same $\beta$ and $L$ ($\beta=(f-L)/f$ in the converging case).

The most important parameter when designing a TERMITES device is the fraction $\beta$ of the unknown beam diameter used to produce the reference. For example, a smaller $\beta$ will make the iterative algorithm more likely to converge to the true physical result since the reference would be from a smaller portion of the unknown beam, and therefore more likely to be free of STCs. But according to the previous discussion, a small $\beta$ will also result in more stringent geometric constraints, potentially making the setup untenable in size or price. There are also limits on $\beta$ since the best signal-to-noise ratio in the computed complex spectrum is when there is perfect fringe contrast. However, since the reference is diverging this requires the beamsplitter in the interferometer to be different than 50:50. As $\beta$ become smaller this becomes a bigger issue and one must eventually compromise with non-ideal fringe contrast.

We now emphasize certain advantages and limitations of the TERMITES technique. Because of the strict requirements on pixel size the device will generally have a very good spatial resolution. This totally eliminates the phase unwrapping ambiguity previously encountered with SEA-TADPOLE and its variants, and will also provide TERMITES with remarkable ability to resolve very fine spatial features, as demonstrated in Ref.\cite{jeandet19}. On the other hand, it also causes the initial data files to be in the tens of Gigabytes. From a physical point of view, the spectrum of the whole beam can only be resolved properly if the reference is as spectrally broad or broader than the unknown beam. So, if the center of the unknown beam where the reference is taken from has a narrower spectrum than the outer portions, then the resulting spectrum after all analysis steps will be narrower than in reality.

The TERMITES technique (and FTS in general) requires taking camera images based on many successive laser shots. Because of this any fluctuations of parameters can have an effect on the final result: intensity fluctuations can be accounted for using a portion of the camera image that is not interfering or by using the integrated signal on the detector, but fluctuations in spectrum, pointing, wavefront, or the STCs on the pulse could have a significant effect. Additionally, since delay steps are not perfect, fluctuations in the delay above a certain level can cause a degradation of the retrieved spectrum~\cite{dorrer00}. Despite this inherent shortcoming of being multi-shot, the few measurements on very large laser systems that tend to have non-trivial fluctuations have been successful~\cite{jeandet19,pariente16}.

\begin{figure}[htb]
	\centering
	\includegraphics[width=83mm]{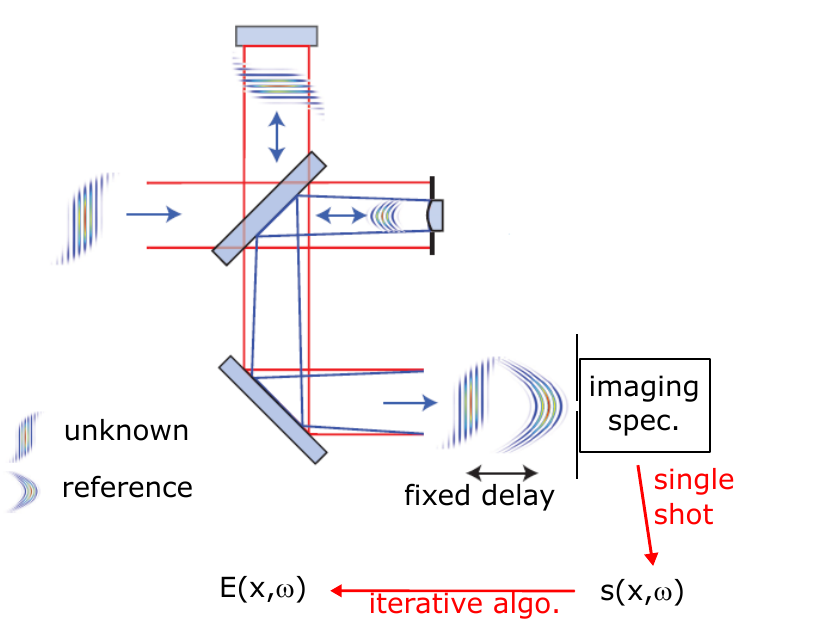}
	\caption{The SEA-TERMITES technique is a single-shot version of TERMITES whereby an imaging spectrometer resolves the cross-spectral density $\hat{s}$ in only one spatial dimension. The idea was proposed in Ref.~\cite{gallet-thesis}, with the visualization style in this figure adopted from Ref.~\cite{pariente16}.}
	\label{fig:TERMITES_2}
\end{figure}

\begin{table*}[tb]
	\begin{tabular}{p{0.21\textwidth}p{0.14\textwidth}p{0.07\textwidth}p{0.12\textwidth}p{0.06\textwidth}p{0.08\textwidth}p{0.14\textwidth}}
		Category & Technique(s) & Spatial dim. & Spectral information obtained by & Single-shot? & Complete? & Type of reference beam\\
		\hline
		\hline
		Extension of established techniques & 2D-SPIDER, SEA-SPIDER, SRSI-ETE & 1D & Spectrometer & Yes & No & Shearing of unknown beam, XPW effect \\
		\hline
		Spatially-resolved spectral-interferometry & SEA-TADPOLE, STARFISH & 2D, usually 1D & Spectrometer & No & Yes, but difficult & Independent or part of unknown beam \\
		\hline
		Spatially-resolved FTS & TERMITES & 2D & FTS & No & Yes & Part of unknown beam \\
		\hline
	\end{tabular}
	\caption{Summary of the techniques discussed in this section, which we deem to be spatially-resolved spectral measurements.}
	\label{tab:table1}
\end{table*}

As a last discussion related to the TERMITES principle, there exists a version of TERMITES that is single-shot, but only resolved in one spatial dimension. We refer to this as SEA-TERMITES~\cite{gallet-thesis}, which is pictured in Fig.~\ref{fig:TERMITES_2}. The concept is simple, rather than scanning over $\tau$ the device can be set up at a single delay and an imaging spectrometer can be installed at the location of the screen or camera. Then a single-shot measurement will produce the TERMITES data at one spatial slice via spectral interferometry. Following this measurement a similar analysis procedure must be followed (except that the data is already resolved in frequency) with vastly decreased data size and therefore speed, with the price of loosing information in one spatial dimension. In the case of TERMITES the success of the iterative algorithm requires that the measured slice is precisely that going through the mutual center of the reference and the unknown beam, and cylindrical symmetry must be assumed. This same equivalence between spectral interferometry and FTS will be seen again briefly in the case of the INSIGHT device, with the same trade-off of being single-shot, but losing one dimension of information and having restrictions on the symmetry of the beam.

As a conclusion of this section, the main properties of the different techniques discussed in this section are summarized in Table~\ref{tab:table1}.

\section{Frequency-resolved spatial measurement techniques}
\label{sec:frequency}

The second major classification of characterization methods corresponds to the techniques that can be referred to as frequency-resolved spatial measurements. This mostly takes the form of techniques that frequency resolve a measurement that is generally used to determine the spatial amplitude and phase profiles of a beam, but will also include more advanced examples that are either similar in philosophy or in methods. This sometimes results in loss of one spatial coordinate, or is made at the cost of resolution.

The measurement of the laser wavefront is important for many experiments involving focused laser beams, especially those operating at high-intensity. Measurement of a laser wavefront generally uses one of the following three approaches: 1- measuring the local slope of the wavefront, e.g. using an array of micro-lenses to create an array of foci whose position depends on the local wavefront (Shack-Hartmann method), 2- interferometry, either externally-referenced or self-referenced (as in the common four-wave lateral shearing interferometry~\cite{Primot:93, chanteloup04}), 3- phase-retrieval algorithms applied to amplitude-only measurements in different $z$ planes ~\cite{10.1117/12.472377}. Essentially, the techniques described in this section are derived from one of these approaches. 

\subsection{Extensions of established wavefront sensing techniques}
\label{sec:wavefront}

What we call here "direct" spatial phase measurements correspond to industry-standard wavefront measurements, which in this section are expanded to include the frequency dimension. We include here both Shack-Hartmann and Four-wave shearing interferometry, although the latter is also related to a following subsection that discusses interferometric methods. Wavefront autocorrelation was already attempted early on~\cite{grunwald03}, where autocorrelation was performed on the sub-foci of an all-reflective Shack-Hartmann device. However, this result was not expanded upon in the literature, so we rather focus on other methods that have been detailed more recently.

\begin{figure}[htb]
	\centering
	\includegraphics[width=83mm]{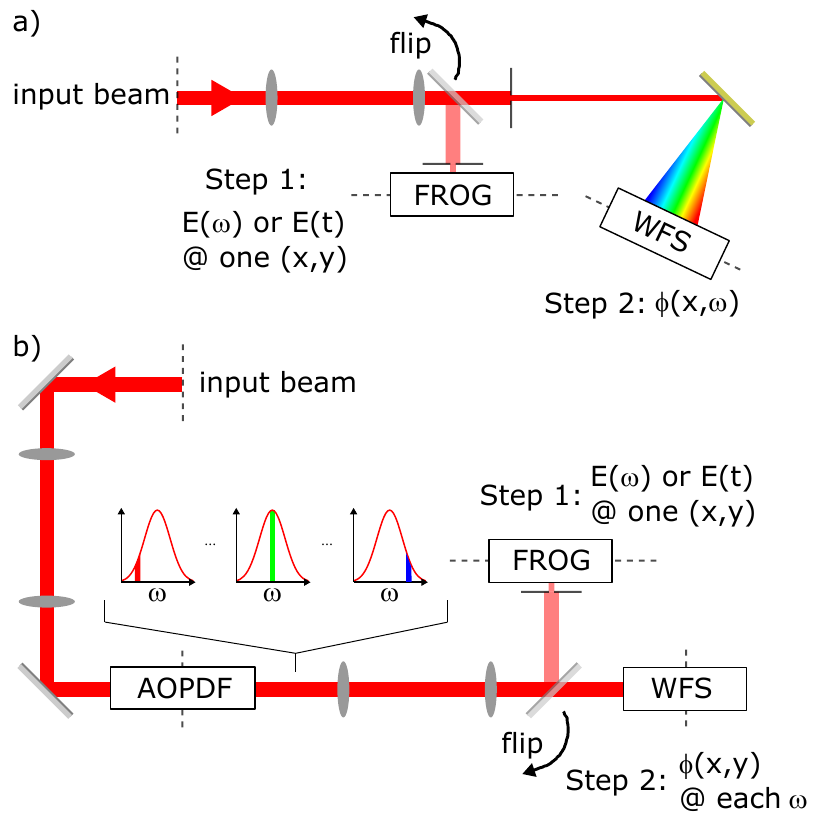}
	\caption{(a) Shackled-FROG technique, based on the schematic from Ref.~\cite{rubino09}. (b) HAMSTER technique, based on the schematic from Ref.~\cite{cousin12}.}
	\label{fig:HAMSTER}
\end{figure}

A first measurement involving a wavefront measurement at a small number of frequencies and propagation calculations~\cite{hauri05} has confirmed that resolving the wavefront spectrally is indeed a valid method for reconstructing or approximating the entire electric field, providing a good foundation for these techniques. A further result resolved in frequency, combined with spectral phase stitching (see section \ref{sec:phase-stitching}), was the Shackled-FROG technique~\cite{rubino09}. This technique, visualized in Fig.~\ref{fig:HAMSTER}(a) is the combination of an imaging spectrometer with a Shack-Hartmann wavefront sensor, as well as a single FROG measurement. Essentially the Shack-Hartmann wavefront sensor is placed at the image plane of an imaging spectrometer, producing an array of sub-foci that are resolved in frequency along the dispersive axis of the grating, and resolved in one spatial dimension. This measurement results in the spatio-spectral phase $\hat{\phi}_S(x,\omega)$ at one sampled $y_0$ position, up to an unknown spatially homogeneous spectral phase. When combined with a FROG measurement, in this case at one point on the same spatial slice $y_0$, the total spatio-spectral phase can be reconstructed. The intensity of the sub-foci can lead to a proxy measurement for the spectral intensity, which means that the spatio-spectral electric field is fully known along one spatial axis.

The HAMSTER technique~\cite{cousin12}, pictured in Fig.~\ref{fig:HAMSTER}(b) also uses a Shack-Hartman device, but keeps both spatial dimensions. This is accomplished by using an acouto-optic dispersive filter (AOPDF) before the wavefront sensor to select narrow regions of the original spectrum. After making multiple such measurements the spatio-spectral phase $\hat{\phi}_S(x,\omega)$ is known, again without pure spectral phase knowledge. A local FROG measurement can lead to the reconstruction of the full spatio-spectral phase, and again the intensity of the sub-foci on the Shack-Hartman device leads to knowledge of the spatio-spectral amplitude. This device then is capable of measuring the full spatio-spectral electric field, but has a few restrictions. In particular, in order to select the narrow spectral regions without adverse effects, the AOPDF must be behind any amplifiers. Because the AOPDF generally has a very small aperture, this limits the type of beams that can be measured without prior demagnification.

\begin{figure*}[tb]
	\centering
	\includegraphics[width=171mm]{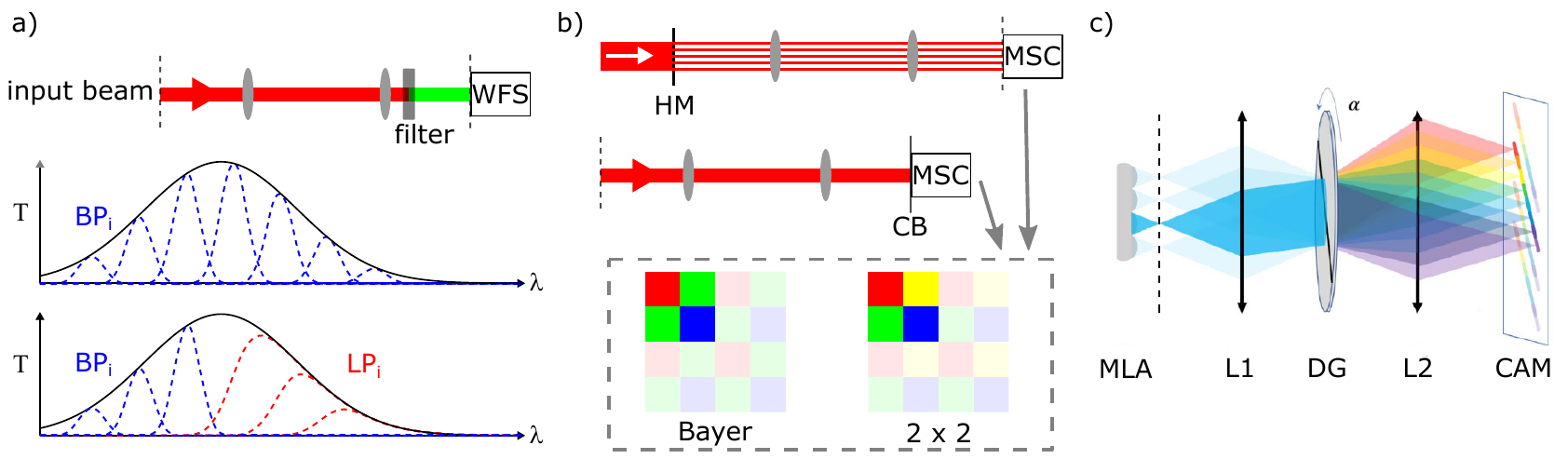}
	\caption{Wavefront measurement possibilities. (a) the simplest example of measuring the wavefront with a standard sensor using various filters for different measurements, where the filters can either being band-pass or a combination of band-pass and high- or low-pass as in Ref.~\cite{kueny18}. (b) The example of using a multi-spectral camera (MSC) to do measurements with either a Hartmann mask (HM) at a Fourier plane or a checkerboard mask (CB) close to the MSC as in Ref.~\cite{dorrer18}. Possible pixel patterns, either Bayer or 2x2 are shown in the inset. (c) Snapshot technique that may be able to resolve spectral data in 3-D using a micro-lens array (MLA) and a tilted diffraction grating (DG). The image in panel (c) is adapted with permission from Ref.~\cite{boniface19} \textcopyright The Optical Society.}
	\label{fig:wavefront}
\end{figure*}

Taking a more direct approach, a Shack-Hartman device has been used in combination with various band-pass and long-pass filters (shown in Fig.~\ref{fig:wavefront}(a)) to assess the wavefront of broadband laser sources, such as a white-light continuum~\cite{hauri05,kueny18} . This method is conceptually similar to the HAMSTER technique, but potentially cheaper to implement and is only limited spatially by the aperture of the sensor and the filters. However, it is difficult to design filters that have a narrow transmission bandwidth, so the wavefront must be constructed via a very small number of frequencies as in Ref.~\cite{hauri05} or a set of measurements that had overlapping spectra as in Ref.\cite{kueny18}. This could lead to errors if the transmission of the filters is not well-known or the spectrum is highly modulated, and will regardless lead to a relatively low spectral resolution. It must also be taken into account that the spectral filters may themselves impart wavefront imperfections. From this point of view the problem of spatio-spectral measurement is somewhat shifted to filter calibration in this technique. There is also a device in development that uses a single filter rotated or translated to shift its transmission~\cite{ranc19}, mitigating some of the mentioned limitations.

Rather than filtering the incoming beam and then measuring with a wavefront sensor, it has also been shown that one can use a multi-spectral camera, combined with either a Hartmann mask or a checkerboard mask (for four-wave lateral shearing interferometry)~\cite{dorrer18} (shown in Fig.~\ref{fig:wavefront}(b)). This approach is elegant and simple, but suffers from a very low spatial resolution, low wavelength resolution, and requires a multi-spectral camera that is generally quite expensive. For example, two common pixel patterns are shown in the inset of Fig.~\ref{fig:wavefront}(b) that only have two or three colors, and reduce the resolution of the CCD by a factor of 4. However, the rapid industrial progress in so-called "snapshot" multi-spectral imaging techniques~\cite{hagen13} could be adapted for the spatio-temporal characterization of ultrashort beams. This application of multi-spectral or hyper-spectral cameras could be transformative.

An example of a snapshot multi-spectral imaging technique was deployed to rapidly characterize a scattering medium~\cite{boniface19} (shown Fig.~\ref{fig:wavefront}(c)) and could be adapted for pulse characterization. This technique used the combination of a lens array and a tilted grating to create an array of sub-foci on a 2D sensor, intimately related to the field of integral field spectroscopy common to astronomy~\cite{allington-smith06}. Because the grating was tilted relative to the axis of the lens array, the dispersed sub-foci do not overlap. With a proper patterning of the lens array, the ability to pack a given spectrum on the 2D sensor can be maximized, although there is still a limitation on the bandwidth of the pulse to be measured. It could be that other methods applied to study scattering media~\cite{liX19}, or the multi-spectral properties of scattering media or multi-mode fibers themselves~\cite{xiong20,xiong19-2,ziv20} could be used to characterize ultrashort pulses, although for now it is highly speculative.

Frequency-resolved detection of the wavefront of the trains of attosecond pulses produced via high-harmonics generation has been done in various implementations~\cite{frumker09,austin11,freisem18,dacasa19}, but we emphasize that the discrete nature of the high-harmonic spectra makes it a significantly different problem than for a single ultrafast pulse with a continuous spectrum.

\subsection{Iterative phase retrieval with frequency resolution}
\label{sec:phase_retrieval}

In this section, we look at measurement methods that rely on iterative phase retrieval algorithms. This type of approach is well known for wavefront measurements of monochromatic beams. The principle is to measure the spatial intensity profile of a beam --which is straightforward to do using a camera-- at multiple $z$ planes separated by known distances. The evolution of this profile as the beam propagates obviously depends on the phase properties of the beam. Iterative algorithms such as the Gerchberg-Saxton one have thus been developed to extract this information from a few measured profiles~\cite{gonsalves79,fienup82,matsuoka00}. Such phase-retrieval methods are already used in concert with deformable mirrors in order to optimize the \textit{on-target} focal spot of high-power lasers~\cite{pharao,oasys,beamtuner}.

However, directly applying this type of approach to a broadband laser beam is doomed to fail if this beam is affected by significant chromatic effects. Indeed, in such a case, the measured intensity profile is the incoherent sum of multiple and potentially different intensity profiles associated to each frequency, $I(x,y)=\sum_{\omega}{I(x,y,\omega)}$, which each evolve in their own way along the propagation axis. Such effects are not taken into account in the iterative phase retrieval algorithm, which will then either poorly converge, or converge to a wrong solution. However, if the intensity profile of the beam is known at each frequency, then this approach can be safely applied independently to all frequencies of the beam. This is the approach discussed in this section, which can be implemented in different ways. 


\begin{figure}[htb]
	\centering
	\includegraphics[width=83mm]{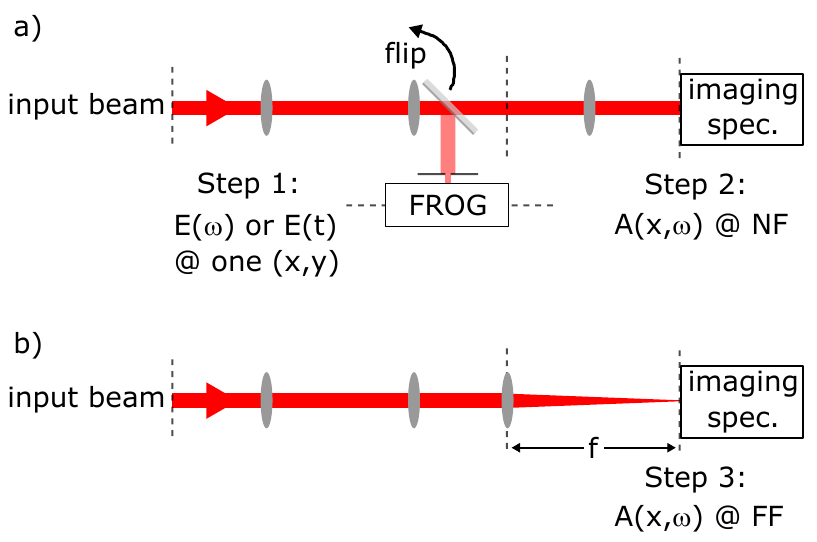}
	\caption{CROAK technique, based on the schematic from Ref.~\cite{bragheri08}, but improved to have less complicated steps.}
	\label{fig:CROAK}
\end{figure}

The first implementation of such a phase-retrieval technique for spatio-temporal measurements was termed the CROAK technique, standing for Complete Retrieval of the Optical Amplitude and phase using the $(k,\omega)$ spectrum~\cite{bragheri08}. This method is detailed in Figure~\ref{fig:CROAK} with a simplified geometry compared to that in the original reference. The technique requires three steps. The first step, in Fig.~\ref{fig:CROAK}(a) is to measure the spectral phase at a well-known position via any method (FROG in this example). The second step, also in Fig.~\ref{fig:CROAK}(a) involves measuring the spatially resolved spectrum of the fundamental beam to be measured in the near-field along one axis using an imaging spectrometer (including the point where the FROG measurement was done). And lastly, the third step requires measuring the spatially resolved spectrum directly in the focus of a lens with a known focal length (and without aberrations) with the same imaging spectrometer, shown in Fig.~\ref{fig:CROAK}(b). The combination of the latter two steps allows for reconstruction of the one-dimensional spatial phase at each frequency using phase retrieval algorithms, within certain limits and requiring certain assumptions. When combined with the spectral phase measurements of the first step this could produce the complete E-field in one spatial dimension and in either frequency or time.

There are many issues with this method, however. Firstly, the measurement of the spatio-spectral amplitude in the near-field and at the far-field must be done on exactly the same slice of the beam, and any aberrations on the beam which lack cylindrical symmetry about the axis of the spectrometer slit could cause significant errors. And secondly, since the spatio-spectral amplitude is measured in the near-field and then at the far-field, there are very tight restrictions on either the size of the near-field beam, the focal length, or the number and size of the pixels in the imaging spectrometer.

\begin{figure*}[tb]
	\centering
	\includegraphics[width=171mm]{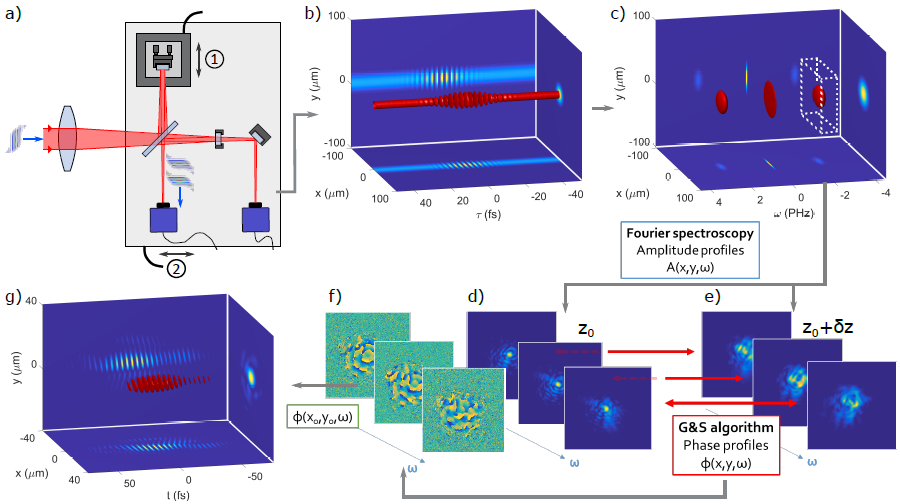}
	\caption{INSIGHT technique. (a) The laser beam to be characterized is focused onto a 2D sensor placed at the output of a simple Michelson or Mach-Zehnder interferometer. A second camera looking at the leakthrough of one end-mirror gives pointing information in order to compensate for any fluctuations during the measurement. A piezoelectric stage in one arm (stage 1 in (a)) is used to scan the delay with sub-optical-period accuracy, and thus measure the spatially-resolved linear autocorrelation of the beam at best focus (longitudinal position $z_0$), shown on (b). The Fourier transform of this signal with respect to delay (c) is then calculated, in order to filter out (white dashed box) the spatially-resolved spectrum. This 3D dataset provides the amplitude profiles at each frequency, shown in (d). This process is repeated at two other longitudinal positions near focus $z_0\pm\delta{z}$ by shifting the interferometer (stage 2 in (a)) in the direction of the incoming beam, thus providing the spectrally-resolved spatial amplitudes at these two other planes (a single panel is displayed in (e)). A Gerchberg Saxton-like phase retrieval algorithm is then applied on these amplitudes at each frequency in order to extract the spatial phase, examples of which are shown in (f). With the addition of an independent measurement of the spectral phase at one position in the beam a Fourier transform with respect to frequency obtains the field in time, the real part of which is shown on (g). This figure is slightly modified with permission from a version in Ref.~\cite{borot18} \textcopyright The Optical Society, and courtesy of A. Borot.}
	\label{fig:INSIGHT}
\end{figure*}

A significant improvement in spatio-temporal measurement using phase retrieval is the INSIGHT technique~\cite{borot18}. Rather than using an imaging spectrometer, the INSIGHT technique resolves the spectral amplitude at each point of the measured beam via spatially-resolved Fourier-transform interferometry around the focus. Knowing the spectrum at each point of the beam, one can obviously determine the spatial profile at each frequency, which is the information required for proper phase-retrieval on broadband beams.

The approach is implemented by splitting the beam near its focus into two copies as shown in Fig.~\ref{fig:INSIGHT}(a), and resolving the interference of these two copies on a standard CCD camera. This has the advantage of resolving the spatial properties in two dimensions. The FTS is performed by taking camera images as the delay $\tau$ is stepped through with sub-cycle accuracy (Fig.~\ref{fig:INSIGHT}(b)), and the spatially-resolved spectrum is calculated via taking the Fourier transform with respect to $\tau$ and selecting only the positive frequencies (Fig.~\ref{fig:INSIGHT}(c)). This procedure removes the large and expensive imaging spectrometer device when compared to CROAK and resolves the second spatial dimension, but requires a delay stage capable of delay steps of a fraction of a wavelength as was the case for the TERMITES technique. 

Performing this temporal scan for one $z$ plane already provides the beam spatio-spectral amplitude. In order to obtain the spatio-spectral phase, the INSIGHT technique requires this spatio-spectral amplitude at multiple planes, just as with the CROAK technique. In order to allow for a better convergence of the phase-retrieval algorithm the FTS is repeated at two additional planes out-of-focus (at $\pm\delta{z}$). Once the spatio-spectral intensity is found around the focus (Fig.~\ref{fig:INSIGHT}(d)) and out-of-focus (Fig.~\ref{fig:INSIGHT}(e)) the phase-retrieval algorithm is done at each frequency to compute the spatio-spectral phase (Fig.~\ref{fig:INSIGHT}(f)). Finally with a single measurement of the spectral phase a one position ($x_0$,$y_0$) the spatio-temporal electric field can be computed (Fig.~\ref{fig:INSIGHT}(g)).

Since INSIGHT is done in-focus it allows for the optics of the measurement device to be very small and lightweight, and also allows the measurement to be done in exactly the location of eventual experiments. The out-of-focus planes are generally measured at $\delta{z}\approx3-10 z_R$, so the camera properties can be optimized to have high-resolution. However, the camera chip should be significantly larger than the beam focus so that high spatial frequencies in the near-field can still be resolved. The INSIGHT device was used successfully for measurements on Terawatt~\cite{borot18} and Petawatt lasers~\cite{jeandet19}. With the addition of a second camera looking at the leak-through of one interferometer arm (shown in Fig.~\ref{fig:INSIGHT}(a)) pointing fluctuations can be measured for each laser shot and numerically corrected, which significantly increases the fidelity of the computed spatially-resolved spectrum~\cite{borot18}. This is especially important for measurement on high-power and low rep-rate systems, where pointing fluctuations can be significant.

A birefringent delay line~\cite{harvey04,brida12} has recently been used for hyperspectral imaging~\cite{perri19} (spatially-resolved Fourier-spectroscopy without phase information), and we have recently implemented this scheme in an INSIGHT device as well~\cite{jolly_prisms}.

Finally, using an imaging spectrometer near the focus of the INSIGHT device could essentially be a single-shot version that is resolved in only one spatial dimension (similar to the single-shot version of TERMITES, SEA-TERMITES), and would be very similar to steps 2 and 3 of the CROAK technique (but done around the focus). However, when compared to SEA-TERMITES, operating very close to the focus would still allow for the optics to be small and lightweight. But we consider the loss of one spatial dimension and the demand for cylindrical symmetry and measurement of the exact same spatial slice to be severe downsides of such a modified version of INSIGHT. If suitable hyperspectral cameras eventually become available, it might be possible to implement a single-shot version of INSIGHT that does not suffer from these limitations, by fitting multiple replicas of the beam into the camera's chip, to measure the spectrally-resolved spatial intensity profile of the beam in multiple $z$ planes in a single-shot. Implementation of this idea, however, is far from straightforward.

\subsection{Interferometric Techniques}
\label{sec:interferometric}

Techniques of this class rely on the interference of the unknown beam with a second beam, considered as a reference, which can either be obtained from the unknown beam itself (self-referenced interferometry), or be an independent perfectly characterized beam. 

A self-referenced interferometric technique commonly used for the spatial characterization of laser beams is spatial shearing interferometry. This is the spatial analogous of the SPIDER technique, where the unknown beam is interfered with a spatially-sheared replica of itself. The resulting interference pattern can be used to determined the spatial derivative of the spatial phase. This technique has been extended for the spatio-spectral characterization of ultrashort laser beams, leading to a technique called "spectrally-resolved spatial-shearing interferometry"~\cite{dorrer02-3}. 

In this technique, a Michelson interferometer is used to generate, on the slit of an imaging spectrometer (oriented along the $x$ axis at the sampling point $y_0$), two beams separated by a delay $\tau$ , with an angle offset represented by $k_x$ and with a spatial shear $X$. The interferogram signal $S$ generated can then be written as follows:

\begin{align}
\label{eq:dorrer}
\begin{split}
&S\left(x,\omega\right)=|\hat{E}\left(x,\omega\right)|^2+|\hat{E}\left(x-X,\omega\right)|^2 \\
&+2|\hat{E}\left(x,\omega\right)\hat{E}\left(x-X,\omega\right)| \\
&\times\cos\left[\hat{\phi}\left(x,\omega\right)-\hat{\phi}\left(x-X,\omega\right)-k_x x-\omega\tau\right] .
\end{split}
\end{align}

\noindent By nulling the shear $X$, it is possible to calibrate for the term $-k_x x-\omega\tau$. After adding the shear again, one has access to the phase difference $\hat{\phi}\left(x,\omega \right)-\hat{\phi}\left(x-X,\omega\right)$, proportional directly to $\partial\hat{\phi}\left(x,\omega\right)/\partial{x}$. From this data, we can reconstruct the phase $\hat{\phi}_S\left(x,\omega\right)=\hat{\phi}\left(x,\omega\right)+\alpha(\omega)$ with $\alpha$ being an arbitrary function of $\omega$. This gives the spatio-spectral phase up to an unknown overall spectral phase, and does not directly give the spectral amplitude. Essentially it is a measurement of the one-dimensional wavefront (i.e. along $x$) resolved in frequency. This combined with a spatially-resolved SPIDER is the 2D-SPIDER technique referenced earlier~\cite{dorrer02-1}.

We note that in principle, by sweeping the position $y$ of the beam on the slit of the spectrometer and then by rotating the beam by 90${}^{\circ}$ around the $z$ axis and by measuring the interferogram at a position $x_0$, the full STC phase $\hat{\phi}_S(x,y,\omega)$ could be obtained (with still an ambiguity of an arbitrary function of $\omega$). However, this is very difficult to do in practice and in fact has not been demonstrated.

As we have seen already, diffraction, including of course the standard linear diffraction grating, is useful for separating the frequencies for doing spatio-spectral measurements. This has so far been to essentially use one dimension of a 2D detector to resolve the frequency while the other dimension remains for one spatial axis (as in SEA-SPIDER~\cite{witting16}, Shackled-FROG~\cite{rubino09}, CROAK~\cite{bragheri08}). However, we saw already briefly in Fig.~\ref{fig:wavefront}(c)~\cite{boniface19} that a cleverly oriented grating can orient the frequency information in such a way that the 2D detector has continuous 2D spatial information along with discrete frequency information. Indeed, this shifts the difficulty from the sensor to the analysis.

\begin{figure}[htb]
	\centering
	\includegraphics[width=84mm]{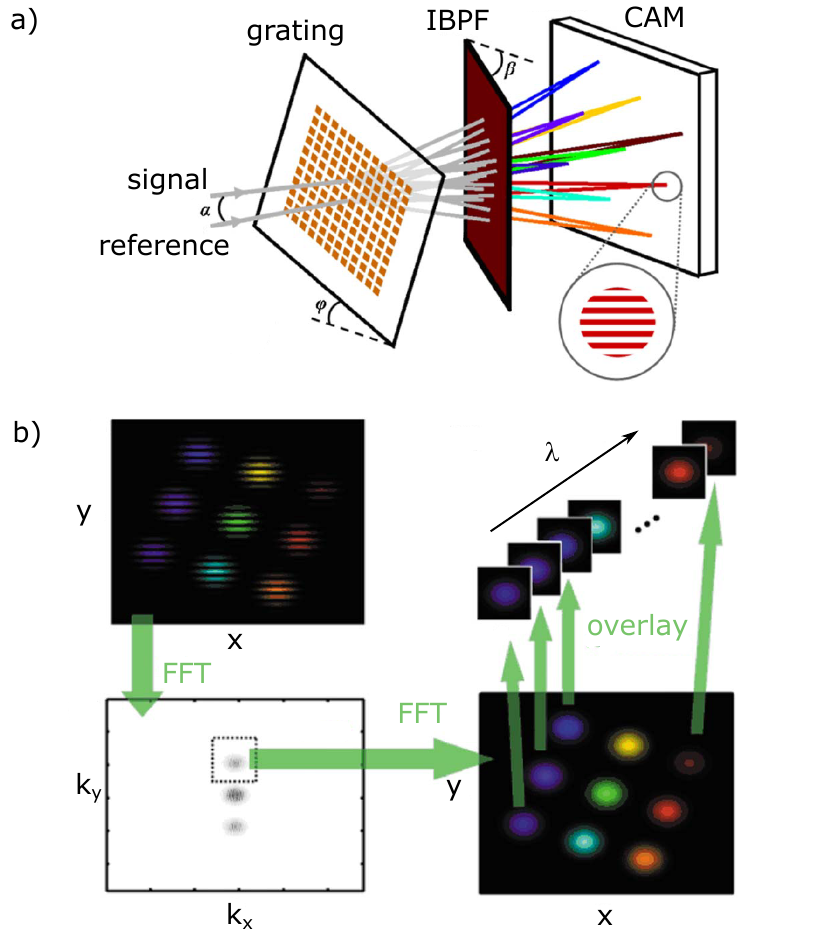}
	\caption{STRIPED-FISH technique. The setup in (a) requires both the unknown beam and a characterized reference, a 2D diffraction grating, an interferometric bandpass filter (IBPF), and a 2D CCD detector. The analysis in (b) involves the standard spatial FFT, but also requires segmenting and overlaying the discrete frequency components. Images modified with permission from Refs.~\cite{gabolde06,gabolde08} \textcopyright The Optical Society.}
	\label{fig:STRIPED-FISH}
\end{figure}

The STRIPED-FISH technique, which is complete and single-shot, uses a tilted 2D diffraction grating and a frequency filter to create spatially-separated profiles at discrete frequencies on a standard 2D CCD chip~\cite{gabolde06,gabolde08,guang14,guang15}. The interference of these dispersed intensity profiles with a \textit{spatio-temporally perfect} reference, or a reference that has been \textit{perfectly characterized in space-time}, in principle allows for measurement of the full spatio-spectral amplitude and phase, and hence for the reconstruction of the full spatio-temporal field. This technique is summarized in Fig.~\ref{fig:STRIPED-FISH}.

\begin{table*}[tb]
	\begin{tabular}{p{0.17\textwidth}p{0.18\textwidth}p{0.11\textwidth}p{0.2\textwidth}p{0.06\textwidth}p{0.08\textwidth}p{0.14\textwidth}}
		Category & Technique(s) & Spatial dim. & Spectral information obtained by & Single-shot? & Complete? \\
		\hline
		\hline
		Extension of established wavefront sensing techniques & HAMSTER, Shackled FROG, hyperspectral wavefront sensors & 1D or 2D & Spectrometer/grating or filter or multispectral camera & Yes & Yes  \\
		\hline
		Spectrally-resolved phase retrieval & CROAK, INSIGHT & 2D (INSIGHT only) & Spectrometer or FTS & Not yet & Yes \\
		\hline
		Interferometry & STRIPED FISH & 2D & Filter & Yes & Yes  \\
		\hline
	\end{tabular}
	\caption{Summary of the main techniques discussed in this section, which we deem to be spectrally-resolved spatial measurements.}
	\label{tab:table2}
\end{table*}

As shown in Fig.~\ref{fig:STRIPED-FISH}(a) the reference and unknown beams are incident with an angle $\alpha$ on a 2D diffraction grating that is tilted by an angle $\varphi$ in a plane perpendicular to the propagation direction of the unknown beam. The grating produces many diffraction orders that are of course diffracted at larger angles from that of the incident beams. An interferometric bandpass filter (IBPF) is placed after the grating, tilted at an angle $\beta$, yet relative to a different plane (see Fig.~\ref{fig:STRIPED-FISH}(a)). Because it is an interferometric filter the transmitted spectrum varies with incidence angle on the filter. Since different diffraction orders of the grating have a different angle of incidence, this causes each order to be filtered to a different bandpass wavelength. The result is a mosaic of spots on the 2D CCD sensor which correspond to the different frequency components of the incident beams. Since the unknown beam and reference beam have a relative angle $\alpha$, this produces spatial interference fringes on each spot, which allows one to extract phase information as well. The analysis steps involve the standard spatial FFT to calculate the amplitude and phase, segmenting the acquired image so that each frequency component can be analyzed independently, and finally stacking the information properly to produce the 3D amplitude and phase (see Fig.~\ref{fig:STRIPED-FISH}(b)).

The strength of STRIPED-FISH is that it is both a complete 3D technique, and single-shot---a significant advantage over techniques like INSIGHT and TERMITES for instance, especially for lasers that have  low repetition-rates or fluctuate from shot to shot. Yet, it has significant drawbacks and limitations. The main limitation is one of principle: it requires a reference beam that covers the full spectrum of the unknown beam, and either has no STCs, or has been fully characterized in space-time. In most real-life cases, this reference needs to be produced from the unknown beam itself, which actually makes STRIPED FISH a self-referenced technique. In practice, this means that this technique is mostly suited to the measurement of the spatio-temporal effects induced by an optical system on a laser beam. A pick-off of this beam prior to this optical system can then be used as a reference, provided this system does not increase or shift the beam spectral content---i.e. this system should be linear. In terms of performance, in order to pack all of the interferograms for all frequencies on the 2D detector, there are strong limits on either the size of the detector, the size of the beam, the number of frequencies, or the spatial resolution. Finally, accurately calibrating the wavelength of each diffraction order, properly stacking the different diffraction orders, and finally producing the correct spatio-spectral phase all require a very careful calibration for each unique device and a very robust analysis algorithm.

The STRIPED-FISH technique has for instance been used to measure the ultrafast lighthouse effect~\cite{guang16} and beams from a multi-mode fiber~\cite{guang17}, where a pick-off on the input beam prior to the optical system under investigation was used as a reference. But due to the previous limitations, this technique has not been in common use so far, despite its conceptual elegance and its complete and single-shot character. In particular, because of the difficulty of producing a reliable reference, it has not yet been used to directly characterize spatio-temporally the output of a complex laser system, to the best of our knowledge.

As a conclusion of this section, the properties of the main techniques discussed in this section are summarized in Table~\ref{tab:table2}.

\section{Analysis and visualization}
\label{sec:analysis_visualization}

The results of a 2-D or 3-D spatio-temporal or spatio-spectral measurement in general consist of a large matrix of complex numbers describing the laser field. For example, in the case of TERMITES or INSIGHT measurements this could be a 3-D complex-valued matrix of a size above 300$\times$300$\times$50 pixels (${x}\times{y}\times\omega$, 4.5 million points). With such a large set of complex data in more than two dimensions, it is not always straightforward to analyze, to extract meaningful physical information, or to visualize the results of a successful measurement. We will outline some methods to analyze the data produced from these measurements with various goals in mind, and in each case will also provide examples of how to effectively visualize the data and the individual analysis steps.

\subsection{Phase-stitching}
\label{sec:phase-stitching}

We start with a post-processing treatment of the measurement data that in practice is often the very first step of the analysis. For many of the techniques described in this tutorial, we have seen that the spatio-spectral phase is actually measured up to an unknown overall spectral phase, that equally applies to all points of the beam. This includes TERMITES, INSIGHT, and direct wavefront measurements with filters or multi-spectral cameras. Therefore these techniques by themselves should be more rigorously called spatio-spectral---rather than spatio-temporal---characterization devices. For other techniques such as SEA-TADPOLE and STRIPED-FISH, only when the technique is done with a suitably characterized reference can the measured spatio-spectral phase be considered complete. All spatio-spectral couplings are still resolved well in every case, and as we will see below, a lot can be said about the beam properties even with this remaining indeterminacy. But without knowledge of the overall spectral phase, the actual spatio-temporal field cannot be calculated.

\begin{figure}[htb]
	\centering
	\includegraphics[width=83mm]{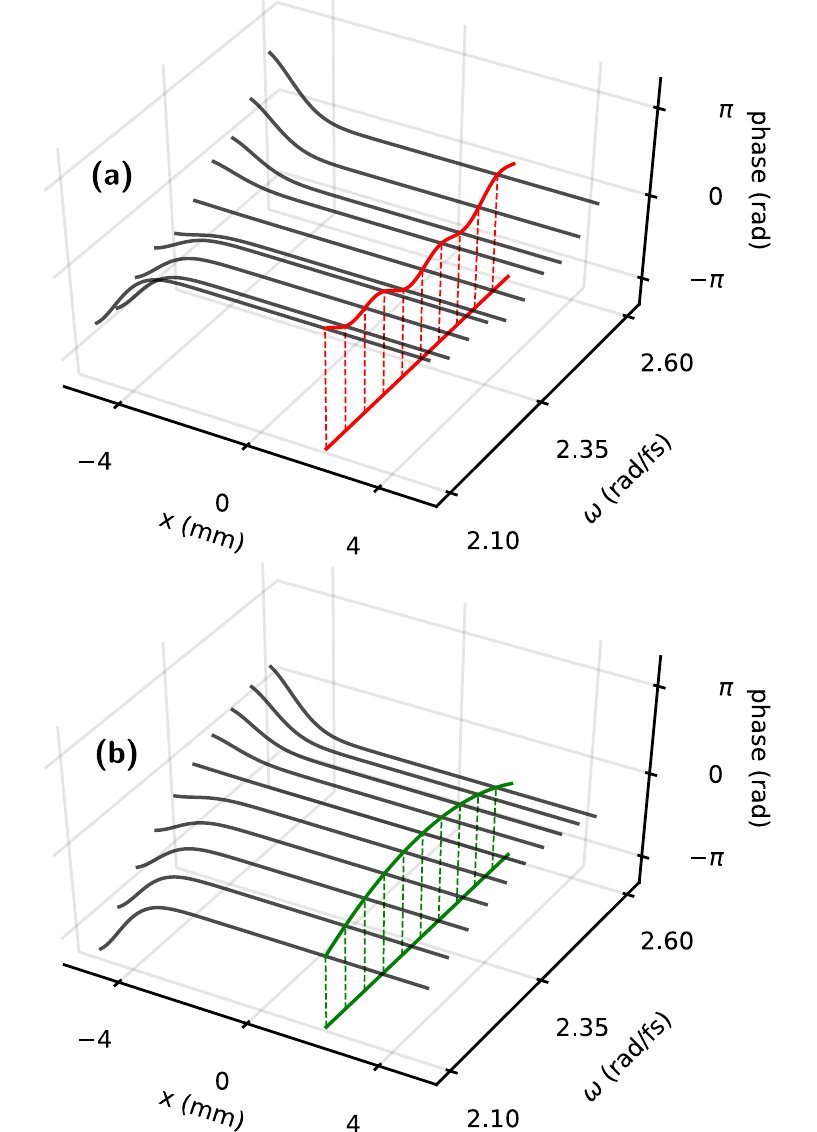}
	\caption{For some measurement techniques, phase stitching is necessary to have the proper phase relationship between frequencies, and be able to calculate the field in the spatio-temporal domain by a Fourier transformation. A measurement technique may be blind to spectral phase as in (a), but produce spatial phase results at many frequencies. A single measurement of spectral phase (b), in this case at $x=2$\,mm, will fix the phase relation and produce the correct spatio-spectral phase at all positions. This figure is courtesy of A. Jeandet.}
	\label{fig:phase-stitching}
\end{figure}

In all cases, a measurement of the spectral phase at a known single point in space can resolve this issue. This single measurement gives the relationship between the retrieved spatial phase maps at different frequencies. Using this measurement one can do "phase-stitching" to transform the data from the spatio-spectral device, with the data at different frequencies being essentially independent of each other, to data having the complete \textit{physical} spatio-spectral phase. This phase-stitching procedure is illustrated in Fig.~\ref{fig:phase-stitching} for 2D data (one spatial coordinate only), but the concept applies equally well for 3D data. Mathematically, this phase stitching operation consists in applying the following transformation to the measured spatio-spectral phase $\hat{\phi}_{meas}(x,y,\omega)$ (displayed in Fig.~\ref{fig:phase-stitching}(a)), to obtain the physical spatio-spectral phase $\hat{\phi}(x,y,\omega)$ (displayed in Fig.~\ref{fig:phase-stitching}(b)):

\begin{equation}
\hat{\phi}(x,y,\omega)=\hat{\phi}_{meas}(x,y,\omega)-\hat{\phi}_{meas}(x_0,y_0,\omega)+\varphi(\omega),
\end{equation}

\noindent where $\varphi(\omega)$ is the spectral phase measured at a given point $(x_0,y_0)$ of the unknown beam (green line in Fig.~\ref{fig:phase-stitching}(b)), using a temporal measurement device such as a SPIDER, FROG, or D-scan for instance. Because performing a local measurement of the spectral phase is much easier on an unfocused beam, this procedure is generally applied to the spatio-spectral phase in the NF. In such a case, when the spatio-spectral measurement is performed in the FF (like in INSIGHT), the measured field needs to be numerically propagated from the FF to the NF before applying phase stitching. 


\subsection{Calculating the magnitude of low-order couplings}
\label{sec:analysis_low}

One of the key steps when analyzing measured data from a spatio-temporal characterization device is generally estimating the magnitude of the lowest-order couplings, or that of the couplings expected to be present based on the nature of the source. This is an essential step because the most common couplings are also typically those of lowest-orders.

Returning to the canonical couplings of AD/PFT and CC/PFC, we can find a straightforward way to calculate the magnitude of these couplings using the phase data of the 3-D matrix. If we consider the NF, these couplings are only concerning the phase, so we reference the reconstructed spatio-spectral phase $\hat{\phi}(x,y,\omega)$ or $\hat{\phi}(r,\omega)$. We can find the AD/PFT via the following relation

\begin{equation}
\gamma_x=\frac{\partial}{\partial\omega}\frac{\partial \hat{\phi}(x,y,\omega)}{\partial x} ,
\end{equation}

\noindent and we can find the CC/PFC via the following relation

\begin{equation}
\alpha=\frac{1}{2}\frac{\partial}{\partial\omega}\frac{\partial^2 \hat{\phi}(r,\omega)}{\partial r^2} .
\end{equation}

\noindent In order to be insensitive to the spectral phase of the measured pulse, the spatial derivate must be performed first. This is important especially for measurements that do not have a pure spectral phase measurement included (i.e. without spectral phase stitching).

\begin{figure}[htb]
	\centering
	\includegraphics[width=83mm]{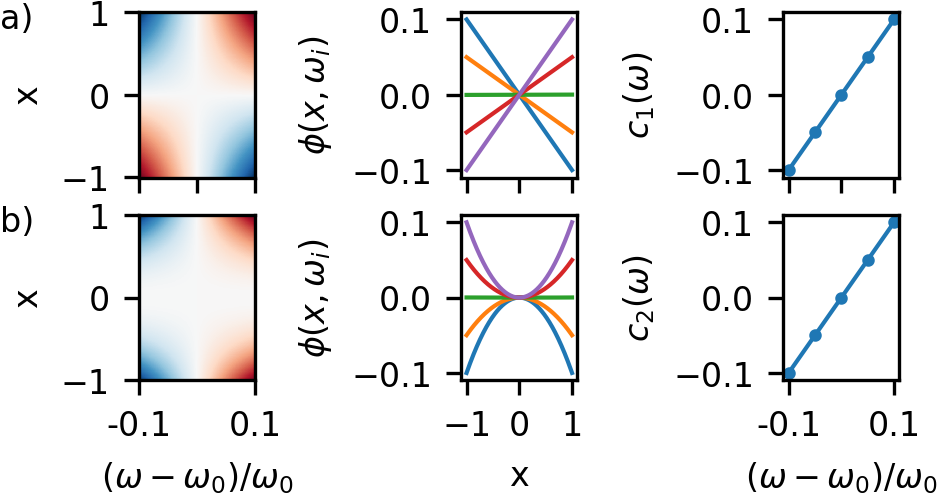}
	\caption{Examples of low-order couplings analysis for (a) AD/PFT and (b) CC/PFC. A slice of the spatio-spectral phase at $y=0$ is shown in the left column. lineouts of this phase at discrete frequencies $\omega_i$ are shown in the central column, which have varying linear or quadratic coefficients for $x$, $c(\omega)$, if the coupling is AD/PFT or CC/PFC respectively. Fitting a linear curve to $c(\omega)$ in each case, as shown in the right column, can result in the magnitude of the coupling. In the case of CC/PFC the behavior is the same in terms of the radial coordinate $r$, but a slice of $x$ at $y=0$ is shown here for simplicity.}
	\label{fig:analysis_1}
\end{figure}

These are simple relationships that make it easy to calculate these STCs when the spatio-spectral phase is a known function, but of course when the phase is represented not as a continuous function, but rather as a discrete data set, derivatives cannot be taken as such. In practice, to find the magnitude of these couplings, the spatial phase at each frequency should rather be fit to a polynomial in space. Then the relevant coefficients at each frequency (either linear in position for AD/PFT or quadratic in radius for CC/PFC) should be fit to a polynomial in frequency. The linear component of this polynomial in frequency is the magnitude of the coupling. More explicitly for AD/PFT:

\begin{equation}
\hat{\phi}(x,y,\omega)=c_1(\omega)\times x ,\quad c_1(\omega)=\gamma_x\times (\omega-\omega_0) ,
\end{equation}

\noindent and for CC/PFC:

\begin{equation}
\hat{\phi}(r,\omega)=c_2(\omega)\times r^2 ,\quad c_2(\omega)=\alpha\times (\omega-\omega_0) ,
\end{equation}

\noindent where in each case $c_i(\omega)$ and the subsequent coupling (either $\gamma$ or $\alpha$) are found via a least-squares regression. Figure~\ref{fig:analysis_1} shows this procedure for both AD/PFT and CC/PFC. It is very important to do these analysis steps on phase data, only within the spectral region where there is significant intensity. Most measurement devices will produce random or highly irregular phase data outside of the real spectral region of the measured beam, which must be ignored because it would negatively influence any fitting.

\subsection{Addressing the magnitude of arbitrary phase couplings}
\label{sec:analysis_high}

Beyond low-order couplings addressed in the previous section, it may be that higher-order phase couplings are expected, or that it is clear there are some effects not of low-order. And besides expectations, in general it can be tedious to fit individual polynomials to 3-D data. Furthermore, when looking at the 3-D data resulting from a spatio-temporal or spatio-spectral measurement, it cannot always be clear whether the high-order aberrations that are present are chromatic or not. Therefore it is necessary to have some type of general way to address this, especially if there is no particular expectation or prediction (which is often the case in the real world). We borrow a standard technique from monochromatic wavefront analysis, and propose to utilize frequency-resolved Zernike polynomials to describe the general phase aberrations present. This was introduced and implemented with great utility in recent work~\cite{borot18,jeandet19}.

The Zernike polynomials are a way to represent an arbitrary function over the unit disk via a set of orthogonal polynomials~\cite{zernike34}. These polynomials can be used to represent the spatial phase of a laser beam over a defined pupil~\cite{born99}, which corresponds to the area where there is significant intensity. Without going into detail, we simply remind that the Zernike polynomials $Z_n^m$ generally have two indices $m$ and $n$ (with $\left|m\right|\le n$) that correspond to the azimuthal and radial degrees of freedom respectively, where $m$ can be negative, but $n$ is limited to the natural numbers. When the phase map is decomposed onto this basis of Zernike polynomials, the result is a list of constants $C_n^m$ corresponding to the amplitude of each polynomial. Algorithmically this is much simpler than fitting arbitrary polynomials to the spatial data, since it amounts to decomposing a known function on a complete orthonormal basis set. We emphasize that these coefficients can be divided by $k=2 \pi /\lambda$, with $\lambda$ the wavelength of the beam under consideration, such that they describe the actual physical distance of displacement/deformation of the wavefront across the beam. Although this is not necessarily the standard practice in wavefront sensing, it is preferable to use this normalization of the coefficients for the analysis of chromatic effects discussed below, and this is what we will assume in the rest of this section.

The extension of the Zernike polynomials to include frequency is quite straightforward. At each frequency $\omega$, the spatial phase map is decomposed on the basis of Zernike polynomials, leading to coefficients $C_n^m(\omega)$ that can now depend on frequency. When a given term does depend on frequency, it can then be concluded that there is a chromatic effect on that Zernike component. For instance, for a beam with PFT, at least one of the coefficients $C_1^{\pm1}$ will vary with frequency - while they would be exactly constant for a tilted beam (provided the normalization mentioned above is used).

\begin{figure*}[tb]
	\centering
	\includegraphics[width=171mm]{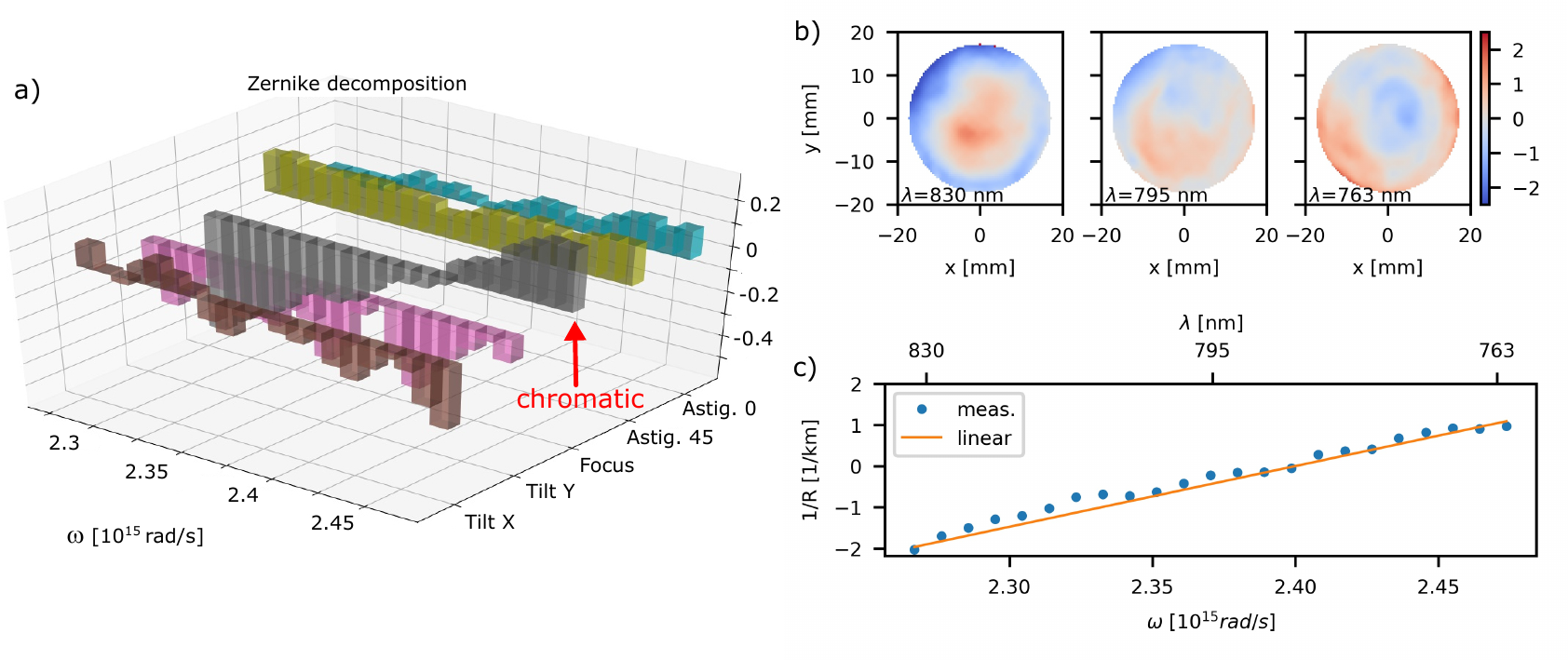}
	\caption{Frequency-resolved decomposition of the spatial phase on Zernike polynomials, and implications. An example of all low-order frequency-resolved aberrations deduced from an INSIGHT characterization of a beam having CC/PFC is displayed in (a). The aberrations besides focus are small but non-zero, and do not have a significant frequency dependence. In contrast, the focus term does have a clear chromaticity, i.e. its value varies with frequency. This variation is linear, which indicates that it indeed corresponds to CC/PFC. Three different phase maps of this beam are shown in (b), spanning the entire spectral width of the laser measured, which show different curvatures represented by different Zernike coefficients in the initial analysis of (a). The magnitude of the CC/PFC in this case can then be deduced from the frequency slope of the wavefront curvature, shown in (c). The data in this figure are adapted from Ref.~\cite{jolly20-1} \textcopyright The Optical Society.}
	\label{fig:Zernike}
\end{figure*}

Such a picture is quite powerful, since it intuitively can show the chromatic nature of different phase aberrations. The intuition of the various aberrations (defocus, astigmatism, etc.) can be utilized to attempt to understand the data that now has the additional dimension of frequency. An example of this data for a beam having mostly CC/PFC in shown in Fig.~\ref{fig:Zernike}(a).

We now explain how to relate these coefficients to different coupling parameters introduced earlier in this tutorial for the canonical couplings AD/PFT and CC/PFT (see section \ref{sec:concepts_manifestations}). For calculating the magnitude of the AD/PFT coefficient $\gamma$, the tilt Zernike terms $C_1^{\pm 1}(\omega)$ first need to be related to the frequency-varying wavefront tilt $\theta$ (in direction $x$ or $y$), through the following relationship:

\begin{equation}
\theta_{x,y}(\omega)\approx\frac{d}{R_p}=\frac{2 C_1^{\pm 1}(\omega)}{ R_p} \label{eq:ZernikeTilt} \\
\end{equation}
\noindent where $R_p$ is the pupil radius used for the Zernike computation, and $d$ is the displacement of the wavefront at a given frequency at the edge of the unit disc defined by the pupil. Note that this relationship assumes that the Zernike modes are normalized to have a modulus of $\pi$ over the unit disc \textbf{(really?)}. The AD/PFT coefficient $\gamma$ is then directly related to the linear slope of $\theta$ via:
\begin{equation}
\gamma_{x,y}=\frac{\omega_0}{c}\frac{\partial\theta_{x,y}}{\partial\omega}\Big|_{\omega_0} \label{eq:ZernikePFT},
\end{equation}

with $\omega_0$ the central frequency of the pulse. 

Following the same reasoning, the CC/PFC coefficient $\alpha$ can be deduced from the frequency-varying Zernike terms for defocus $C_2^0(\omega)$, by relating both quantities to the frequency-resolved wavefront curvature $1/R(\omega)$:
%

\begin{align}
\frac{1}{R(\omega)}&=\frac{2d}{d^2 + R_p^2}\approx\frac{2d}{R_p^2}=\frac{4\sqrt{3} C_2^0}{ R_p^2} \label{eq:ZernikeDefocus} \\
\alpha&=\frac{\omega_0}{2c}\frac{\partial(1/R(\omega))}{\partial\omega}\Big|_{\omega_0} \label{eq:ZernikePFC},
\end{align}

\noindent where $d$ is the same as before. A schematic of this is shown in Fig.~\ref{fig:Zernike}(b)--(c). The phase maps at three frequencies in Fig.~\ref{fig:Zernike}(b) show qualitatively the varying curvature, but the linear fit to the frequency-dependence of $1/R(\omega)$ in Fig.~\ref{fig:Zernike}(c) produces the clear quantitative value of the CC/PFC based on Eq.~(\ref{eq:ZernikePFC}).


Both of these examples are essentially identical to the straightforward approach outlined in the previous section for low-order couplings, but require a different set of steps. Depending on one's priorities and capabilities either method should result in the same quantitative result. The obvious advantages of the Zernike polynomial method are that the higher-order aberrations are generated for free, a similar analysis can be done for the chromatic nature of these higher-order aberrations, and all of the technical considerations of fitting at low or high orders are not relevant. However, it is important to realize that choosing an appropriate pupil for the Zernike calculations is very important for calculating the correct result. 

\subsection{Assessing the total effect of couplings}
\label{sec:analysis_total}

As a complementary step to quantifying the magnitude of specific couplings, which have identified causes or effects, it is useful to quantify the overall impact of all couplings present. There are methods to quantify the various effects of phase and amplitude couplings separately or together, and again they depend on the application. In some optical setups, for example a NOPA~\cite{harth18} or a multi-pass cell for pulse compression~\cite{weitenberg17,lavenu18}, there may be a significant effect on the homogeneity of the spectral amplitude when the system is not properly aligned. So in these cases a quantity can be used to assess this level of homogeneity when optimizing. This assessment of spectral homogeneity over the spectrum can be defined for example by the spectral overlap integral $V$~\cite{weitenberg17,lavenu18}

\begin{equation}
V(r)=\frac{\left[\int\sqrt{I(\lambda,r)\times I(\lambda,r=0)}d\lambda\right]^2}{\left(\int I(\lambda,r)d\lambda\right)\times\left(\int I(\lambda,r=0)d\lambda\right)}.
\end{equation}

\begin{figure}[htb]
	\centering
	\includegraphics[width=83mm]{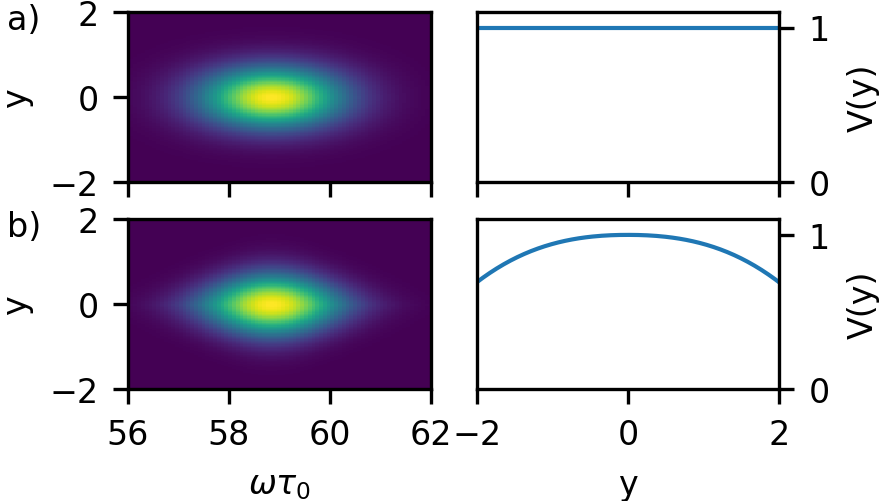}
	\caption{A perfect pulse (a), with the spectral amplitude shown on the left and the calculated (perfect) overlap integral shown on the right. An example pulse with spectral amplitude on the left in (b) having a decreasing spectral width with increasing $y$, with the spectral overlap integral calculated the right, showing a decrease away from the axis.}
	\label{fig:analysis_3}
\end{figure}

\noindent This integral is essentially comparing the spectrum at off-axis positions to the spectrum on-axis. As the spectrum becomes more homogeneous this integral will approach 1 at every transverse position. An example of this calculation is shown in Fig.~\ref{fig:analysis_3}.

Of course in many applications the phase is important, and in general the phase on the NF beam (which is more often analyzed) will have a larger effect than the amplitude on desired parameters, such as in-focus pulse duration or peak intensity. So in addition to looking at the spectral amplitude, various spatio-temporal Strehl ratios can quantify the impact of spatio-temporal phase distortions~\cite{pariente16,jeandet19}. The beam measured has a spatio-temporal or spatio-spectral amplitude $A(x,y,\omega)$ and phase $\hat{\phi}(x,y,\omega)$. The commonly known Strehl ratio ($\textrm{SR}_\textrm{WFS}$, associated with standard wavefront sensors) quantifies the effect of the frequency-averaged wavefront on the focusing of the frequency-averaged beam profile. This is usually performed on data that is already averaged (via measurement on a CCD camera), i.e. $\textrm{SR}_\textrm{WFS}=I\left[\overline{A}(x,y)e^{i\overline{\phi}(x,y)}\right]/I\left[\overline{A}(x,y)\right]$, where the upper bar symbol indicates an average over frequency. We use $I[ ]$ to denote the calculation of the focused intensity of a given beam. With knowledge of the full 3D intensity and phase, more nuanced versions of this quantity can be calculated as we now show.

Although there are many possible definitions of a spatio-temporal Strehl ratio, we will focus on only a few versions to demonstrate the concept, which have been used in the previous works~\cite{pariente16,jeandet19}. The Strehl ratio assessing the impact of all phase distortions both chromatic and not, termed $\textrm{SR}_\textrm{Full}$, compares the fully measured beam with a beam having zero phase at every frequency

\begin{equation}
\textrm{SR}_\textrm{Full}=\frac{I\left[\hat{A}(x,y,\omega)e^{i\hat{\phi}(x,y,\omega)}\right]}{I\left[A(x,y,\omega)\right]}.
\end{equation}

\noindent When $\textrm{SR}_\textrm{Full}$ is less than one, it represents the departure in focused intensity from the perfect case of the fully measured beam. It should be the representation of the physically existing pulse intensity. Note that with this definition, the value of $\textrm{SR}_\textrm{Full}$ also depends on the spectral phase of the beam -which not the case with the usual, spatial-only, definition of the Strehl ratio of laser beams: indeed, a chirped laser pulse, even without any spatio-temporal coupling (i.e. the chirp is spatially homogeneous), necessarily has $\textrm{SR}_\textrm{Full}<1$.

The Strehl ratio assessing the impact of only the chromatic phase distortions, which we call $\textrm{SR}_\textrm{STC}$, compares the measured beam with the frequency-averaged wavefront subtracted to a beam having zero phase at each frequency

\begin{equation}
\textrm{SR}_\textrm{STC}=\frac{I\left[\hat{A}(x,y,\omega)e^{i\left(\hat{\phi}(x,y,\omega)-\overline{\phi}(x,y)\right)}\right]}{I\left[A(x,y,\omega)\right]}.
\end{equation}

\begin{figure}[htb]
	\centering
	\includegraphics[width=83mm]{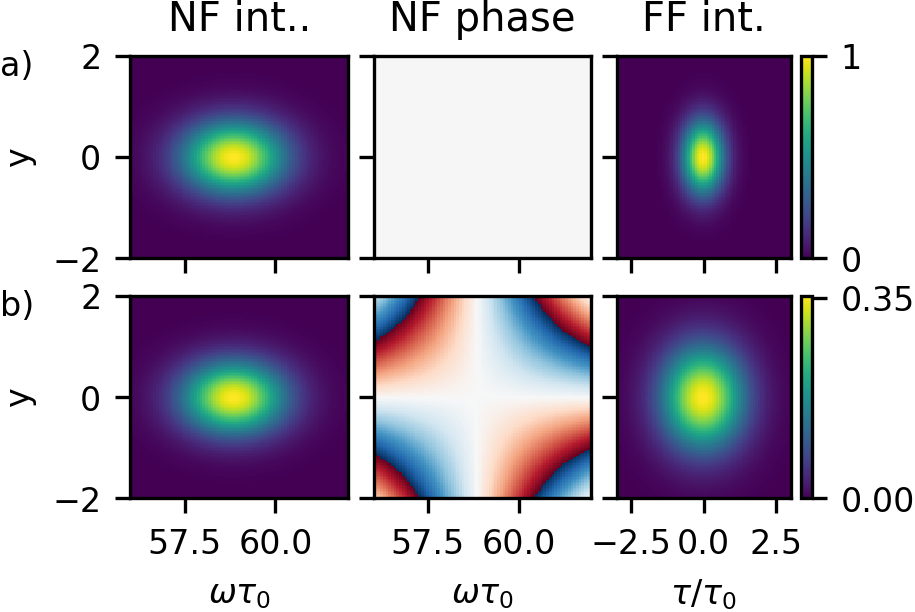}
	\caption{Visualization of the Strehl ratio calculations for one simple case. In (a) a perfect beam, having flat phase, is defined to have an intensity of 1 in the far-field. In (b) a beam with AD/PFT in the nearfield produces a beam with a lower intensity in the far-field, which would correspond to a calculation of $\textrm{SR}_\textrm{Full}=\textrm{SR}_\textrm{STC}=0.35$.}
	\label{fig:analysis_4}
\end{figure}

\noindent The physical case that $\textrm{SR}_\textrm{STC}$ describes is that where a deformable mirror was implemented perfectly so as to remove all non-chromatic wavefront distortions, but all chromatic aberrations still remain. This is useful if it is known that achromatic aberrations exist (and are either impossible to remove or not necessary to remove at that moment), and one wants to assess the impact of STCs only. A simple example is shown in Fig.~\ref{fig:analysis_4}.

Note that if there are no chromatic phase distortions then $\textrm{SR}_\textrm{Full}=\textrm{SR}_\textrm{STC}$. These definitions of $\textrm{SR}_\textrm{Full}$ and $\textrm{SR}_\textrm{STC}$ were used on measurements of the BELLA PW system in recent work~\cite{jeandet19}.

Lastly, If one wants to quantify the effect of all spatio-temporal distortions, both in phase and amplitude, then a mixed ratio can be calculated as was done in~\cite{pariente16}. In that case it was calculated by comparing the measured beam to a beam with zero phase in space and frequency, and also with the amplitude replaced by the average in space and the average in frequency, i.e. $\textrm{SR}_\textrm{mixed}=I\left[\hat{A}(x,y,\omega)e^{i\left(\hat{\phi}(x,y,\omega)\right)}\right]/I\left[\overline{A}(x,y)\overline{A}(\omega)\right]$. However, we believe that looking at the impact of phase and amplitude effects separately generally provides more insight.

\subsection{Visualization}
\label{sec:visualization}

Due to the complexity and multi-dimensional nature of spatio-temporal couplings, visualization is an important issue~\cite{rhodes17,li18-1}. Even after a successful measurement using one of the devices described, it is not trivial to properly discern the couplings present, nor is it simple to properly communicate the magnitude of the couplings. Therefore visualization is crucial to both assess initial measurements to guide the analysis priorities, but also to communicate the impact after analysis has taken place.

\begin{figure*}[tb]
	\centering
	\includegraphics[width=150mm]{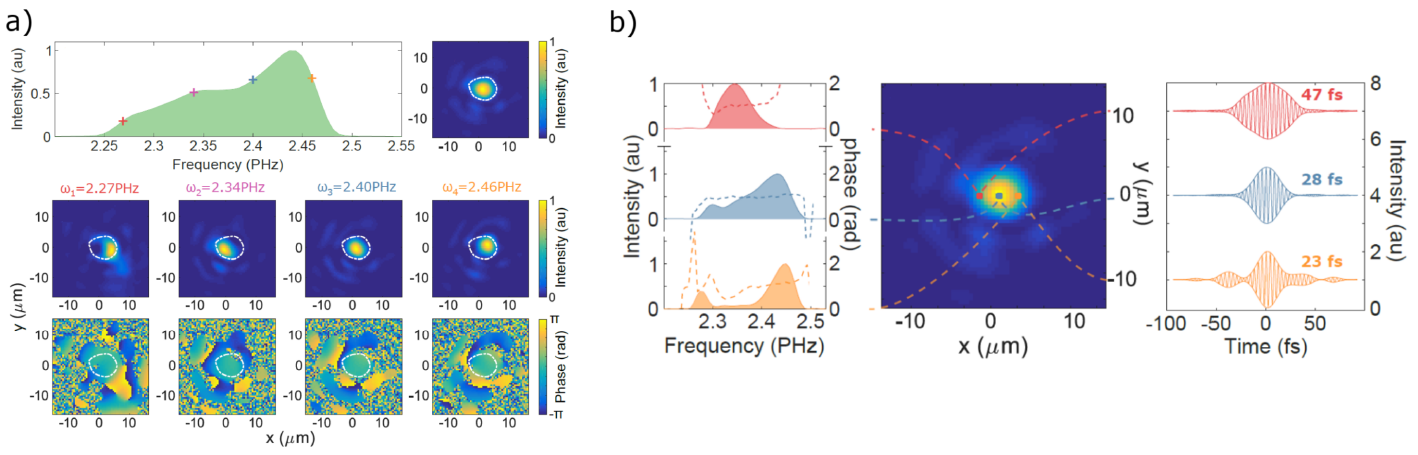}
	\caption{Visualization options for the output of spatio-temporal or spatio-spectral measurements. These are (a) spatial measurements visualized at discrete frequencies, and (b) spectral and/or temporal properties visualized at discrete spatial coordinates. An additional view was already shown in Fig.~\ref{fig:Zernike}(a), the 3D view of the frequency-resolved Zernike coefficients. (a) and (b) are shown in the far-field, but can just as easily be done in the near-field. All subfigures are reproduced with permission from Ref.~\cite{borot18} \textcopyright The Optical Society.}
	\label{fig:visualization}
\end{figure*}

We showcase a few examples of visualization options in Figure~\ref{fig:visualization} (taken from Ref.~\cite{borot18}). These methods are: spatial properties visualized at discrete frequencies (Fig.~\ref{fig:visualization}(a)), spectral and/or temporal properties visualized at discrete spatial coordinates (Fig.~\ref{fig:visualization}(b)) and, as already discussed, the frequency-resolved Zernike coefficients visualized in a 3-D format (Fig.~\ref{fig:Zernike}(a)).

The method used for visualization depends strongly on the desired knowledge. If one is applying a certain spatio-temporal coupling in order to induce a given mechanism at the experimental focus, then visualizations exactly as in Fig.~\ref{fig:visualization}(a) or (b) are likely most helpful. This is because they give a direct visualization of certain properties where they are important. However, if one is a laser physicist looking to remove undesired spatio-temporal couplings, then the same views as in Fig.~\ref{fig:visualization}(a) or (b) may be desired, but on the collimated beam rather than at the focus. This is because the bulk of most laser amplifiers and the optics that may induce undesired couplings act on the collimated beam (although likely at increasing diameter throughout a laser chain). Because of this the views in the near-field may provide more direct input into the source of undesired couplings.

The last important view was already shown in Fig.~\ref{fig:Zernike}(a), the 3D view of the frequency-resolved Zernike polynomials, is likely useful to all scientists. This is because, although the Zernike coefficients are calculated based on the near-field beam, their nature also provides direct input in to the manifestation of any given chromatic effect in the focus. For this reason the frequency-resolved Zernike coefficients may be the most universal and helpful view of all. Since they show no amplitude information, other views will always be necessary as a complement. The data shown in Fig.~\ref{fig:Zernike}(a) is for a beam with CC/PFC (i.e. a linear slope in frequency of the focus) to contrast with that already shown in Ref.~\cite{borot18}.

There are more compact methods to visualize spatio-temporal couplings. One example is where rather than a color scale representing the amplitude or intensity, it corresponds to the local instantaneous frequency. This is especially relevant for couplings where the spatio-spectral amplitude changes with propagation. A systematic review of visualization using this method along with many examples was presented in Ref.~\cite{rhodes17}. Additionally, three dimensional stationary views of pulse intensity can be made with quadrants cut out (see many examples in Ref.~\cite{li18-1}) or with constant intensity contours (see for example Fig.~7 of Ref.~\cite{borot18}).

Beyond the stationary views discussed, it is often useful and instructive to use movies to achieve multiple objectives. This includes: 1) panning or rotating a fixed 3-D plot in order to have a more immediate sense of the 3-D presentation (as in Supplementary movies 1 and 3 of Ref.~\cite{pariente16}, or Visualization 2 of Ref.~\cite{borot18}), 2) showing a 2-D map and stepping through a third parameter (time or frequency) as the movie progresses (as in Supplementary movie 2 of Ref.~\cite{pariente16}), or movies 1 and 2 of Ref.~\cite{jeandet19}), or 3) visualizing a 2-D or 3-D property as the beam is numerically propagated through space where the movie steps through time or propagation distance (as in Visualization 1 of Ref.~\cite{borot18}, or movie 3 of Ref.~\cite{jeandet19}). Although movies are not necessarily as useful as stationary plots within scientific journal articles, they are becoming better integrated in certain journals and their use is becoming more prevalent. Moreover, movies are an extremely useful tool for analysis for a scientist when interpreting results, so for the reader of this tutorial they could be important.

\section{Conclusion}
\label{sec:conclusion}

In this tutorial we have introduced spatio-temporal couplings in a detailed fashion and reviewed techniques ranging from the simple to the complex for characterizing ultrashort laser pulses completely. This included very simple qualitative techniques, established temporal characterization methods extended to include one or all spatial dimensions, and advanced methods using a variety of techniques. The fact that this work is a tutorial was especially stressed in the order of introducing techniques and the level of detail included for a small number of them. From this point of view, it should not be treated as a full review of STC characterization (of which there is a good recent example~\cite{dorrer19}).

In addition to some past results or techniques that have not been discussed, there are many up-and-coming techniques which may prove to be integral to making spatio-temporal characterization more widespread in the community. For example, the STRIPED-FISH technique is the only technique employed for ultrashort pulses that is single-shot, although it requires a reference. A reference-free single-shot method that is more simple to implement experimentally is the grand challenge of this field. Indeed, techniques such as TERMITES and INSIGHT are functioning well and on the road to becoming available products for the community, but they are still methods that require scanning over many independent pulses. It may be that intuition from the mature but separate world of hyperspectral imaging~\cite{hagen13} may provide innovation for pulse characterization if they are improved to handle the broad spectrum of ultrashort pulses, or even via the field of imaging through scattering media~\cite{boniface19, liX19}.

The methods exposed in this paper for visible and near-infrared can be used as inspiration for characterization of sources in other wavelength ranges. This includes the much shorter wavelengths in attosecond pulse (see Ref.~\cite{dacasa19}) and the longer wavelengths of a growing number of mid- and far-infrared ultrafast sources. This is important because, for example, attosecond pulses generated from gases are considered to often have extreme levels of spatio-temporal couplings depending on the precise generation parameters~\cite{wikmark19}. The chief difficulty in developing devices for these exotic wavelengths is generally the components: optics such as mirrors, beamsplitters, and filters are commonplace for near-infrared sources, but can be quite bulky, expensive, or perform worse for extreme wavelengths. Even beyond sources of different wavelengths, ultrafast vector beams --- beams with a spatially-varying polarization --- add a completely new challenge to characterization. There are some solutions in development~\cite{alonso19}, and this will surely become a very active area.

Most of the examples of either simulated or real STCs in this tutorial were simple in nature, mostly in order to clearly demonstrate the concept. These simple STCs have many applications as discussed in the introduction. However, there are many exotic STCs or exotic scenarios where STCs can be an avenue for fine control of physical mechanisms. These mechanisms include: Simultaneous space-and-time focusing caused by focusing a beam with spatial chirp in the nearfield~\cite{zhu05,durfee12,heF14}; Spatio-temporal light springs~\cite{pariente15}, relevant potentially for laser-plasma acceleration~\cite{vieira18}, and extended to the attosecond regime~\cite{porras19-3}; A "Flying Focus" in the focus of a beam with longitudinal chromatism and temporal chirp~\cite{sainte-marie17,froula18,jolly20-1} for Raman amplification~\cite{turnbull18-1}, ionization waves of arbitrary velocity~\cite{palastro18,turnbull18-2}, or photon acceleration~\cite{howard19}; spatial chirp or chromatic focusing for high harmonic generation in gases~\cite{hernandez-garcia16,holgado17}; steering of beams in laser-plasma acceleration due to pulse-front tilt~\cite{popp10,thevenet19-2,mittelberger19} and the effect on the polarization of betatron radiation~\cite{schnell13}; circumventing intrinsic limits of laser-plasma acceleration~\cite{debus19}; pulse-front tilt for dielectric laser acceleration~\cite{plettner08,wei17}; THz beams with tilted pulse-front for traveling-wave electron acceleration~\cite{walsh17}; In-band noise filtering of high-power lasers~\cite{wangJ18}; Diffraction-free space-time wave packets~\cite{kondakci16, kondakci17, kondakci19-1, bhaduri19-1, bhaduri19-2, kondakci19-2, yessenov19-1, yessenov19-2}, among many others.

The recent activity in designing new spatio-temporal characterization devices and the wealth of applications in ultrafast physics underscores the importance of the field. With many Terawatt and Petawatt lasers coming online across the world~\cite{danson19}, and pulses with few-cycle duration becoming ever more commonplace, the increase in familiarity with the concepts in this tutorial is paramount for the community to successfully utilize these sources and to characterize and troublehoot their spatio-temporal performance.

\section*{Acknowledgements}

We acknowledge Antoine Jeandet for general discussions and for supplying the phase-stiching figure.

\appendix
\section{calculation of the spatio-spectral phase of beams with pulse-front distortions}
\label{sec:appendixA}

We consider a beam of central frequency $\omega_0$ whose spatio-temporal field is described by a function of the following form:
\begin{equation}
\nonumber
E(x,t)=f(x) g[t-t_0(x)] e^{i \omega_0 t}
\end{equation}
This corresponds to a beam whose temporal profile $g(t)$ is invariant in space, but whose arrival time $t_0(x)$ depends on the position $x$ in the beam---while the wave front at the carrier frequency, described by the last term of the equation, is assumed to be flat and normal to the $z$ axis. The term $f(x)$ describes the spatial envelope of the beam, and will be omitted in all following calculations, as it appears as an overall factor in all equations. For simplicity, we restrict the analysis to one transverse spatial dimension only, but it can easily be generalized to two transverse coordinates.

The spatio-spectral description of this beam is obtained by performing a Fourier-transformation with respect to $t$. To carry out this transformation, we first rewrite the previous equation as:

\begin{equation}
\nonumber
E(x,t)=[g(t) \otimes \delta[t-t_0(x)] ] \times e^{i \omega_0 t},
\end{equation}

\noindent where $\otimes$ is the symbol for the convolution product. The Fourier-transform of this field is then given by:

\begin{align} 
\nonumber
\hat{E}(x,\omega)&=[\hat{g}(\omega) \times FT\left\{\delta[t-t_0(x)]\right\} ] \otimes FT\left\{e^{i \omega_0 t}\right\} \\
&=[\hat{g}(\omega) \times e^{i\omega t_0(x)}]  \otimes \delta(\omega-\omega_0) \nonumber\\
&=e^{i\delta\omega \: t_0(x)}\hat{g}(\delta \omega), \nonumber
\end{align}

\noindent where $\hat{g}(\omega)$ is the Fourier-transform of $g(t)$, and $\delta \omega=\omega-\omega_0$ is the frequency offset from the central frequency $\omega_0$. 

This equation shows that a pure pulse front distortion $t_0(x)$ in the time domain  is entirely described in the spectral domain by a spatio-spectral phase $\hat{\phi}(x,\omega)=\delta \omega \: t_0(x)$. In the case of pulse front tilt, we have $t_0(x)=\gamma x$ leading to $\hat{\phi}(x,\omega)=\gamma \: \delta \omega \: x$. In the case of pulse front curvature, $t_0(x)=\alpha x^2$ leading to $\hat{\phi}(x,\omega)=\alpha \:\delta \omega \: x^2$. These expressions of the spatio-spectral phase are the ones discussed in section~\ref{sec:concepts_manifestations} of the main text. We now provide a more detailed discussion of these two cases, which is actually very useful to understand some of the subtleties of STCs and their metrology. 

We first analyze the mathematical differences between a beam with PFT, and a perfect STC-free beam propagating at an angle with the $z$ axis. In the later case, the field writes:

\begin{align}
\nonumber
E(x,t)&=g(t-\gamma x) e^{i \omega_0 (t-\gamma x)} \\
&=[g(t)e^{i \omega_0 t}]\otimes \delta(t-\gamma x). \nonumber
\end{align}

\noindent When going to the spectral domain, a calculation similar to the previous one leads to:

\begin{equation}
\hat{E}(x,\omega)= e^{i \gamma x \omega}\hat{g}(\delta \omega). \nonumber
\end{equation}

\noindent This shows that in the spatio-spectral domain, the only difference between a beam with PFT, and a perfect tilted beam, lies in a subtle difference in the spatio spectral phase: in the former case, $\hat{\phi}(x,\omega)=\gamma \delta \omega x$, while in the later, $\hat{\phi}(x,\omega)=\gamma \omega x$. This has important consequences for STC metrology: a measurement method can only distinguish these two types of beams, and hence detect PFT, if it can differentiate these two types of spatio-spectral phases.

A similar analysis shows that in the spatio-spectral domain, a perfect STC-free curved beam (e.g. a perfect beam just after a perfect focusing optic) is described by the spatio-spectral phase $\hat{\phi}(x,\omega)=\alpha \omega x^2$. This again only slightly differs from the case of a beam with PFC, where $\hat{\phi}(x,\omega)=\alpha \delta \omega x^2$. 

This last case actually leads to an interesting question. Let us consider again a perfect beam without STC, which goes through a perfect focusing optic. Just after this optic, the field has the form:

\begin{equation}
E(x,t)=g(t-\alpha x^2) e^{i \omega_0 (t-\alpha x^2)}. \nonumber
\end{equation}

\noindent The beam wave front and pulse front are both curved by the same amount. This form of field is not separable as a product of a function of time and a function of space. According to the definition of section~\ref{sec:concepts_general} (Eq.~(\ref{eq:STC})), this would imply that such a beam presents STC. Yet, intuitively, one would not consider this beam as suffering from STC---or equivalently chromatic aberrations. 

The point of view of the authors is that this apparent contradiction is a weakness or flaw of the present commonly-used definition of STC, which will have to be clarified by further theoretical work. One potential solution would be to consider that a beam has no STC when there exists a transformation $t'=h(t,x)$ such that $E(x,t')$ can be decomposed as $E(x,t')=f(x)g(t')$. With such a definition, a perfect curved beam would then be free of STC, since the transformation $t'=t-\alpha x^2$ makes it separable. Whether this definition makes sense in a more general case remains on open question. 

\section*{References}

\bibliographystyle{unsrt}
\bibliography{biblo_tutorial}
\end{document}